\documentclass[pre,aps,onecolumn,superscriptaddress,showpacs]{revtex4}
\usepackage{amsmath,amssymb,graphicx}
\usepackage{epsfig}
\usepackage{graphicx}
\usepackage{textcomp}

\begin{document}

\title{Extreme Fluctuations of Current in the Symmetric Simple Exclusion Process: a Non-Stationary Setting}
\author{Arkady Vilenkin}
\email{vilenkin@mail.huji.ac.il}
\affiliation{Racah Institute of Physics, Hebrew University of
Jerusalem, Jerusalem 91904, Israel}
\author{Baruch Meerson}
\email{meerson@mail.huji.ac.il}
\affiliation{Racah Institute of Physics,  Hebrew University of
Jerusalem, Jerusalem 91904, Israel}
\author{Pavel V. Sasorov}
\email{pavel.sasorov@gmail.com}
\affiliation{Keldysh Institute of Applied Mathematics, Moscow 125047, Russia}



\begin{abstract}
\noindent
We use the macroscopic fluctuation theory (MFT) to evaluate the probability distribution ${\cal P}$ of extreme values of integrated current  $J$ at a specified time $t=T$ in the symmetric simple exclusion process (SSEP) on an infinite line.
As shown recently [Phys. Rev. E \textbf{89}, 010101(R) (2014)], the SSEP belongs to the elliptic universality class. Here, for very large currents, the diffusion terms of the MFT equations can be neglected compared with the terms coming from the shot noise. Using the hodograph transformation and an additional change of variables, we reduce the ``inviscid" MFT equations to Laplace's equation in an extended space. This opens the way to an exact solution. Here we solve the extreme-current problem for a flat deterministic initial density profile with an arbitrary density $0<n_0<1$. The solution yields the most probable density history of the system conditional on the extreme current, $J/\sqrt{T}\to \infty$, and leads to a super-Gaussian extreme-current statistics, $\ln {\cal P}\simeq -\Phi(n_0) J^3/T$, in agreement with  Derrida and Gerschenfeld [J. Stat. Phys. \textbf{137}, 978 (2009)]. We calculate the function $\Phi(n_0)$ analytically. It is symmetric with respect to the half-filling density $n_0=1/2$, diverges at $n_0\to 0$ and $n_0 \to 1$ and exhibits a singularity $\Phi(n_0) \sim |n_0-1/2|$ at the half-filling density $n_0=1/2$.

\end{abstract}
\maketitle

\tableofcontents
\nopagebreak

\section{Introduction}
\label{intro}
Large fluctuations of current flowing through physical systems far from thermodynamic equilibrium have become a central subject of nonequilibrium statistical mechanics.  Stochastic lattice gas models \cite{Spohn,KL99,L99}, which capture
different aspects of matter and energy transport in a simple schematic form, have been extensively used for studying large current fluctuations.   A popular model is the symmetric simple exclusion process (SSEP) \cite{Spohn,KL99,L99,SZ95,
S00,D07,BE07,KRB10} which describes unbiased random transport of particles and accounts, in a simple way, for  inter-particle repulsion. In the SSEP there can be at most one particle on a lattice site. A particle can randomly hop to each of the neighboring sites with equal probability if that site is unoccupied. If it is occupied, the move is disallowed. This model describes full counting statistics of mesoscopic conductors \cite{Levitov,Buttiker,Jordan1,Jordan2,D07}, and is also relevant in
a host of transport problems in materials science, cell biology, and biophysics \cite{Chou}.

Another extensively studied unbiased lattice gas model is the Kipnis--Marchioro–-Presutti (KMP) model of stochastic heat flow \cite{KMP}. It involves a lattice of static agents (``oscillators") which randomly redistribute energy among neighbors. The KMP model was suggested to mimic heat conduction in a crystal. By virtue of the \textit{a priori} stochastic dynamics, the Fourier's law of heat conduction for the KMP model was established rigorously  \cite{KMP}.

A convenient coarse-grained description of diffusive lattice gases, including the SSEP and the KMP model,  is provided by a Langevin equation \cite{Spohn,KL99}. In one spatial dimension, this equation reads
\begin{equation}
\label{Lang}
     \partial_t n = \partial_{x} [D(n)\, \partial_x n] +\partial_x \left[\sqrt{\sigma(n)} \,\eta(x,t)\right],
\end{equation}
where $n(x,t)$ is the particle number density, $\eta(x,t)$ is a zero-mean Gaussian noise which is delta-correlated both in space and in time:
\begin{equation}
\left\langle \eta(x,t)\eta(x_1,t_1)\right\rangle=\delta(x-x_1)\, \delta(t-t_1),
\label{deltacorr}
\end{equation}
and the brackets denote ensemble averaging. As one can see from Eq.~(\ref{Lang}), a diffusive lattice gas is fully characterized, at large distances and long times, by the diffusion coefficient $D(n)$ and additional coefficient, $\sigma(n)$. The latter coefficient comes from the shot noise of particle transport and is equal to twice the mobility of the gas \cite{Spohn}. The functions $D(n)$ and $\sigma(n)$ obey the Einstein relation: $F^{\prime\prime}(n)=2 D(n)/\sigma(n)$, where $F(n)$ is the equilibrium free energy of the homogeneous lattice gas with density $n$ \cite{Spohn}, and primes denote derivatives with respect to the argument. For the SSEP one has $D=1$ and $\sigma(n)=2n(1-n)$.
For the KMP model $D=1$ and $\sigma(n)=2n^2$.

Most of the work on large current fluctuations in diffusive lattice gases dealt with large but finite systems driven by two different reservoirs of particles or heat baths. These studies brought about a new level of understanding of large fluctuations of \emph{nonequilibrium steady states}, see Refs. \cite{D07,Jona} for reviews. An alternative, non-stationary setting, where a major progress has been achieved, involves finite (periodic) undriven systems and deals with statistics of integrated current at a specified (and sufficiently long) time \cite{D07,HEPG}. Here we will be dealing with still another non-stationary setting. It assumes an infinite system and addresses statistics of current passing through the origin at a specified time \cite{DG2009a,DG2009b,KM_var,varadhan,MS2013,MS2014}. Large current fluctuations in this regime are still poorly understood.   Here the integrated current statistics depends on time and on the initial density profile $n(x,t=0)$ . A convenient initial condition is step-like,
\begin{equation}
\label{step}
n(x,t=0) =
\begin{cases}
n_-,   & x<0,\\
n_+,  & x>0,
\end{cases}
\end{equation}
or flat in the particular case of $n_-=n_+=n_0$. Integrated current $J$ -- the total number of
particles (or the total energy) passing into the half-line $x>0$ during a given time $T$ -- is specified by the condition
\begin{equation}
\label{current0}
\int_0^\infty dx\, [n(x,T)-n_{+}] =J\,.
\end{equation}
In this non-stationary setting one needs to be careful in defining the averaging procedure \cite{DG2009b}. In the \emph{quenched}
setting the initial density profile (\ref{step}) is deterministic, and one only has to average over different realizations of the stochastic process. In the \emph{annealed} setting one allows equilibrium fluctuations in the initial condition (\ref{step}) and averages over them as well. In other words,  the initial density profile at $x<0$ ($x>0$) is chosen from the equilibrium probability distribution corresponding to density $n_-$ (correspondingly, $n_+$). As a result, the most probable initial density profile, conditional on a specified integrated current in the annealed setting, is different from a step function \cite{DG2009b}.

Let us return to Eq.~(\ref{Lang}). With the shot noise term neglected it reduces to
the deterministic diffusion equation
\begin{equation}\label{diffusion}
\partial_t n = \partial_x \!\left[D(n) \,\partial_x n\right].
\end{equation}
The solution of Eq.~(\ref{diffusion}) with the initial condition \eqref{step} yields the \emph{average} integrated current at time $T$. The actual current $J$ fluctuates around this average, and the main quantity of interest is the probability density ${\cal P}(J,T,n_-,n_+)$ which exhibits dynamic scaling behavior at large $T$ \cite{DG2009a,DG2009b,KM_var}:
\begin{equation}\label{scaling1}
\ln {\cal P}(J,T;n_-,n_+) \simeq -\sqrt{T} \,s(j, n_{-}, n_{+}),\;\;\;j=J/\sqrt{T},
\end{equation}
and $\sqrt{T}$ is the characteristic diffusion length scale.  Alternatively, one can work with the moment generating function of $J$ which also exhibits scaling behavior at large $T$:
\begin{equation}
   \left\langle e^{\lambda J} \right\rangle = \int dJ\, e^{\lambda J} {\cal P}(J,T; n_-,n_+) \sim  \int dj \, e^{\sqrt{T}\left[\lambda j-s(j,n_-,n_+)\right]} \sim e^{\sqrt{T}\,\mu(\lambda,n_-,n_+)}, \label{GF}
\end{equation}
where
\begin{equation}\label{GF1}
 \mu(\lambda,n_-,n_+)= \max_{j}\, [\lambda j-s(j,n_-,n_+)] .
\end{equation}
Derrida and Gerschenfeld \cite{DG2009a} obtained an exact solution of the current statistics problem for the microscopic SSEP in the annealed setting.  Taking the long time limit of their exact result, they found $\mu(\lambda,n_-,n_+)$ in this case. They also determined $\mu(\lambda,n_-,n_+)$ for the KMP model, again in the annealed setting  \cite{DG2009b}, by establishing a connection between the SSEP and KMP model
in the annealed setting. This connection appears at the level of \emph{macroscopic fluctuation theory} (MFT) of Bertini, De Sole, Gabrielli, Jona-Lasinio, and Landim \cite{Bertini,Jona}. The MFT is well suited for studying large deviations of different quantities in diffusive lattice gases, as it employs in a smart way a natural small parameter of the problem: the typical noise strength. The latter scales as $1/\sqrt{N}\ll 1$, where $N$ is the typical number of particles in the relevant region of space which becomes large at sufficiently large $T$. The MFT can be formulated as a classical Hamiltonian field theory \cite{Bertini,DG2009b,Tailleur}, and we will use the Hamiltonian language in the following.

The present work deals with the statistics of integrated current in the quenched setting.  An MFT formulation of this problem was obtained in Ref. \cite{DG2009b}, and we will recap it shortly. However, the MFT equations are hard to solve analytically. As of present, the large deviation functions $s(j,n_-,n_+)$ and $\mu(j,n_-,n_+)$, entering Eqs.~(\ref{scaling1}) and (\ref{GF}) for the quenched setting, are only known exactly for non-interacting random walkers \cite{DG2009b}, where $D=1$ and $\sigma(n)=2n$. In the absence of a complete solution for models of interacting particles one can probe different asymptotic regimes where perturbative treatments can be developed.
Krapivsky and Meerson \cite{KM_var} calculated, for diffusive lattice gases with $D(q)=1$ and arbitrary $\sigma(q)$, the asymptotics of $s(j,n_-,n_+)$ and $\mu(j,n_-,n_+)$ when $j$ is close to the rescaled average current $\langle j \rangle =\langle J(T) \rangle/\sqrt{T}$. These asymptotics describe small Gaussian fluctuations of the integrated current around the mean. For the SSEP in the quenched setting
the asymptotic of $s(j,n_-,n_+)$ \cite{KM_var} is
\begin{equation}\label{small}
    s(j,n_-,n_+)\simeq \frac{(j-\langle j \rangle)^2}{2 V},
\end{equation}
where $\sqrt{\pi}\, \langle j \rangle =n_--n_+$ and
\begin{equation}\label{asymfullquen}
\sqrt{2\pi}\,V =n_+ + n_- - \frac{(n_+ +n_-)^2}{2} - \frac{3-2\sqrt{2}}{2}(n_+ -n_-)^2.
\end{equation}
In particular, for $n_-=n_+=n_0$ one obtains $\sqrt{2\pi} \,V=\sigma(n_0)=2n_0(1-n_0)$ \cite{KM_var,varadhan}.

Of great interest is the opposite regime of extremely large currents, $j\to \infty$ \cite{DG2009b,varadhan,MS2013,MS2014}. For the non-interacting random walkers the large-$j$ asymptotic of the exact expression for $s$ is super-Gaussian in $j$ \cite{DG2009b}:
\begin{equation}\label{PRW}
 s(j,n_-,n_+)\simeq \frac{j^3}{12 n_{-}^2}.
\end{equation}
Derrida and Gerschenfeld \cite{DG2009b} conjectured that the super-Gaussian decay $s \sim j^3$  holds for a whole class of \emph{interacting} gases, and proved this conjecture for lattice gases with $D=\text{const}$ and $\sigma(n)\leq n+\text{const}$ for $0\leq n\leq n_{\text{max}}$, and $\sigma(n)=0$ otherwise \cite{varadhan1}.

The next step in the analysis of the extreme current statistics was made in Ref. \cite{MS2014} which identified two different universality classes of diffusive lattice gases with respect to this statistics -- the elliptic and hyperbolic classes. These classes are determined by the sign of the second derivative $\sigma^{\prime\prime}(n)$.  For the elliptic class, $\sigma^{\prime\prime}(n)<0$, the Derrida--Gerschenfeld conjecture $s \sim j^3$ holds \cite{MS2014}, as the large deviation function $s(j,n_-,n_+)$ behaves as
\begin{equation}\label{Pgeneric}
s(j,n_-,n_+)\simeq f(n_{-},n_+) j^3
\end{equation}
with an a priori unknown $f(n_{-},n_+)$. Ref. \cite{MS2014} put forward a ``road map" towards finding this function analytically for the SSEP [actually, for any diffusive lattice gas with $D(n)$ bounded from above and
$\sigma(n)=a n -bn^2$, where $a>0$ and $b>0$]. This progress was possible because, at large $j$, the MFT equations for the elliptic class of gases can be
simplified by neglecting the diffusion terms, see below.  The resulting ``inviscid" equations can be transformed into hydrodynamic equations for effective inviscid compressible fluid with a negative pressure, ensuing an elliptic flow \cite{MS2014}. This effective hydrodynamics is exactly soluble, at least in principle, via the hodograph transformation \cite{LLfluidmech,Courant}. The solution yields the super-Gaussian statistics (\ref{Pgeneric}) and describes the optimal path of the system -- the most probable density history conditional on a given extreme value of integrated current. The optimal mode of transfer of extreme current here involves a large-scale inviscid flow
and includes static and traveling discontinuities \cite{MS2014}.

For systems of the hyperbolic class, $\sigma^{\prime\prime}(n)>0$, the situation is very different, as shown in \cite{MS2013} on the example of the KMP model. Here the optimal mode of transferring an extreme current is a short propagating energy density pulse. A proper description of this pulse demands an account of the diffusion terms in the MFT equations. Furthermore, the dominant contribution to $s(j,n_-,n_+)$ comes from the pulse itself, rather than from the large inviscid flow regions. As a result, the probability of observing an extreme value of current is much higher here than what is predicted by the Derrida-Gerschenfeld scaling, $-\ln {\cal P}/\sqrt{T} \sim j^3$. In particular, for the KMP model one obtains a sub-Gaussian extreme current statistics, $-\ln {\cal P}/\sqrt{T} \sim j \ln j$ \cite{MS2013}.

Let us return to the models of elliptic class, exemplified by the SSEP. The only specific example, explicitly solved for extreme currents in Ref. \cite{MS2014}, dealt with a flat density profile $n_-=n_+=1/2$  (the quenched setting) at $t=0$. As this initial condition respects the particle-hole symmetry of the SSEP, the hodograph solution greatly simplifies. In this paper we extend the extreme-current analysis of Ref. \cite{MS2014}. We still consider a flat initial density profile, $n_-=n_+=n_0$, but allow $n_0$ to take any value between $0$ and $1$. We obtain a complete analytic solution for the optimal path of the system and determine the function $f(n_0,n_0) \equiv \Phi(n_0)$.  These results are presented, in a pictorial way, in Figures \ref{qvofx} and \ref{Phifig}, respectively.

In section \ref{recap} we recap the MFT formulation \cite{DG2009b,KM_var} of the large-current statistics problem for the SSEP and give an overview of the flow structure as described by the inviscid limit of the MFT equation. In section \ref{hodo} we perform the hodograph transformation of the inviscid MFT equations, and then an additional transformation which maps the problem into a Dirichlet problem for Laplace's equation in an extended hodograph space. Section \ref{solution} exposes the solution of the Dirichlet problem and the ensuing hodograph solution. In the same section we calculate the function $\Phi(n_0)$ and discuss its properties. In section \ref{solhopf} we obtain solutions in non-hodographic regions. We briefly discuss our results in Section \ref{discus}.

\section{Macroscopic fluctuation theory, inviscid limit and flow character}
\label{recap}
Here we recap the MFT formulation \cite{DG2009b} of the large-current statistics problem for the SSEP, and its inviscid limit \cite{MS2014}.   Rescaling  $t$ and $x$ by $T$ and $\sqrt{T}$, respectively, we can rewrite Eq.~(\ref{current0}) as
\begin{equation}
\label{current1}
\int_0^\infty dx\, [q(x,t=1)-n_0] =j.
\end{equation}
We can assume $j>0$ without loss of generality. The particle number density field $q(x,t)$ and the canonically conjugate ``momentum" density field $p(x,t)$ obey Hamilton equations \cite{Bertini,DG2009b,Tailleur}
\begin{eqnarray}
\partial_t q &=& \partial_x^2 q
-  \partial_x \left[\sigma(q)\, \partial_x p\right],\label{q:eqfull}\\
\partial_t p &=& - \partial_{x}^2 p
- \frac{1}{2}\sigma^{\prime}(q)\!\left(\partial_x p\right)^2,\label{p:eqfull}
\end{eqnarray}
where $\sigma(q)=2q(1-q)$. Equations~(\ref{q:eqfull}) and (\ref{p:eqfull}) can be obtained as saddle-point equations of the field theory corresponding to the Langevin equation (\ref{Lang}). The Hamiltonian functional is $H=\int_{-\infty}^\infty dx\,h$, where
\begin{equation}\label{hath}
 h = -\partial_x q\, \partial_x p
+(1/2) \,\sigma(q)\!\left(\partial_x p\right)^2.
\end{equation}
Once $q(x,t)$ and $p(x,t)$ are determined, one can calculate the mechanical
action $s=\int\int dt dx \left(p\,\partial_t q - h\right)$, which reduces to
\begin{eqnarray}
  s  &=& \frac{1}{2}\int_0^1 dt \int_{-\infty}^\infty dx \,\sigma(q) (\partial_x p)^2, \label{action0}
\end{eqnarray}
and yields the large deviation function $s(j,n_0,n_0)$ from Eq.~(\ref{scaling1})  \cite{Bertini,Tailleur,DG2009b,KM_var}. For the quenched setting we are interested in, the boundary condition for $q(x,t)$ at $t=0$ is given by the equation
\begin{equation}
\label{stepq}
q(x,t=0) =n_0,
\end{equation}
see  Eq.~\eqref{step} with $n_-=n_+=n_0$.  Additional boundary condition, at rescaled time $t=1$, comes from the minimization of $s$ under the constraint~(\ref{current1}) \cite{DG2009b}:
\begin{equation}
\label{p_step}
p(x,t=1) = \Lambda \,\theta(x),
\end{equation}
where $\theta(x)$ is the Heaviside step function. The
Lagrange multiplier
$\Lambda>0$ is fixed by Eq.~(\ref{current0}).

For extreme currents, $j \to \infty$, one can neglect the diffusion terms in Eqs.~(\ref{q:eqfull}) and (\ref{p:eqfull})  and arrive at the \emph{inviscid} MFT equations \cite{MS2014}:
\begin{eqnarray}
\partial_t q +  \partial_x \left[\sigma(q)\, v \right]&=&0,\label{d1}\\
\partial_t v + \frac{1}{2}\partial_x \left[\sigma^{\prime}(q) v^2\right]&=&0,\label{d2}
\end{eqnarray}
where we have differentiated Eq.~(\ref{p:eqfull}) with respect to $x$ and introduced  the momentum density gradient $v(x,t)=\partial_x p(x,t)$. The inviscid Hamiltonian is
\begin{equation}
\label{Ham}
H_0=\int_{-\infty}^{\infty} dx \,h_0, \;\;\text{where}\;\;h_0=\frac{1}{2}\,\sigma(q) v^2 ,
\end{equation}
whereas, by virtue of $H_0=const$,  the action (\ref{action0}) becomes
\begin{eqnarray}
  s  = \frac{1}{2} \int_0^1 dt \int_{-\infty}^\infty dx \,\sigma(q) v^2 = \int_0^1 dt\,H_0=H_0. \label{action1}
\end{eqnarray}
The boundary condition (\ref{p_step}) now reads
\begin{equation}\label{delta}
    v(x,t=1) =\Lambda \,\delta(x),
\end{equation}
where $\delta(x)$ is the Dirac's delta function.

The inviscid MFT problem has three local conservation laws.  The local conservations of the $q$ and $v$ fields are evident from Eqs. (\ref{d1}) and (\ref{d2}). In fact, they hold already in the full, unreduced MFT formulation, see Eq. ~ (\ref{q:eqfull}) and the $x$-derivative of Eq.~(\ref{p:eqfull}). The unreduced MFT Hamiltonian density $h$ from Eq.~(\ref{hath}) is only conserved globally. However, its inviscid counterpart $h_0$ from  Eq.~(\ref{Ham}) is conserved locally \cite{MS2014}, as it evolves by the continuity equation
\begin{equation}\label{inviscid10}
    \partial_t h_0 +\partial_x (h_0 u) = 0
\end{equation}
with the effective velocity $u=\sigma^{\prime}(q) v$.

The inviscid MFT formulation is consistent with the Derrida-Gerschenfeld conjecture (\ref{Pgeneric}).
Indeed, the inviscid MFT equations (\ref{d1}) and (\ref{d2})
are invariant under the transformation $x/\sqrt{\Lambda} \to x$
and $v/\sqrt{\Lambda} \to v$. Under this transformation $s$ from Eq.~(\ref{action1}) becomes $\Lambda^{3/2} s_1$,
where $s_1$ is the action obtained when the boundary condition
(\ref{delta}) is replaced by the condition $v(x,1)=\delta(x)$. In its turn, the transformed Eq.~(\ref{current1}) is
$\int_0^{\infty}dx\,[q(x,1)-n_0]=j/\sqrt{\Lambda}$, therefore $j=\sqrt{\Lambda} j_1$,
where $j_1$ is the integrated current obtained with the boundary condition $v(x,1)=\delta(x)$.
This immediately leads to Eq.~(\ref{Pgeneric}) with $f(n_0,n_0)\equiv \Phi(n_0)=s_1/j_1^3$ \cite{MS2014}. Furthermore,
\begin{equation}\label{fany}
\Phi(n_0) = \frac{s_*}{j_*^3},
\end{equation}
where $s_*=s_*(n_0)$ and $j_*=j_*(n_0)$ are the action and the integrated current for \emph{any} chosen parametrization of the problem.

\begin{figure}
\includegraphics[scale=0.4] {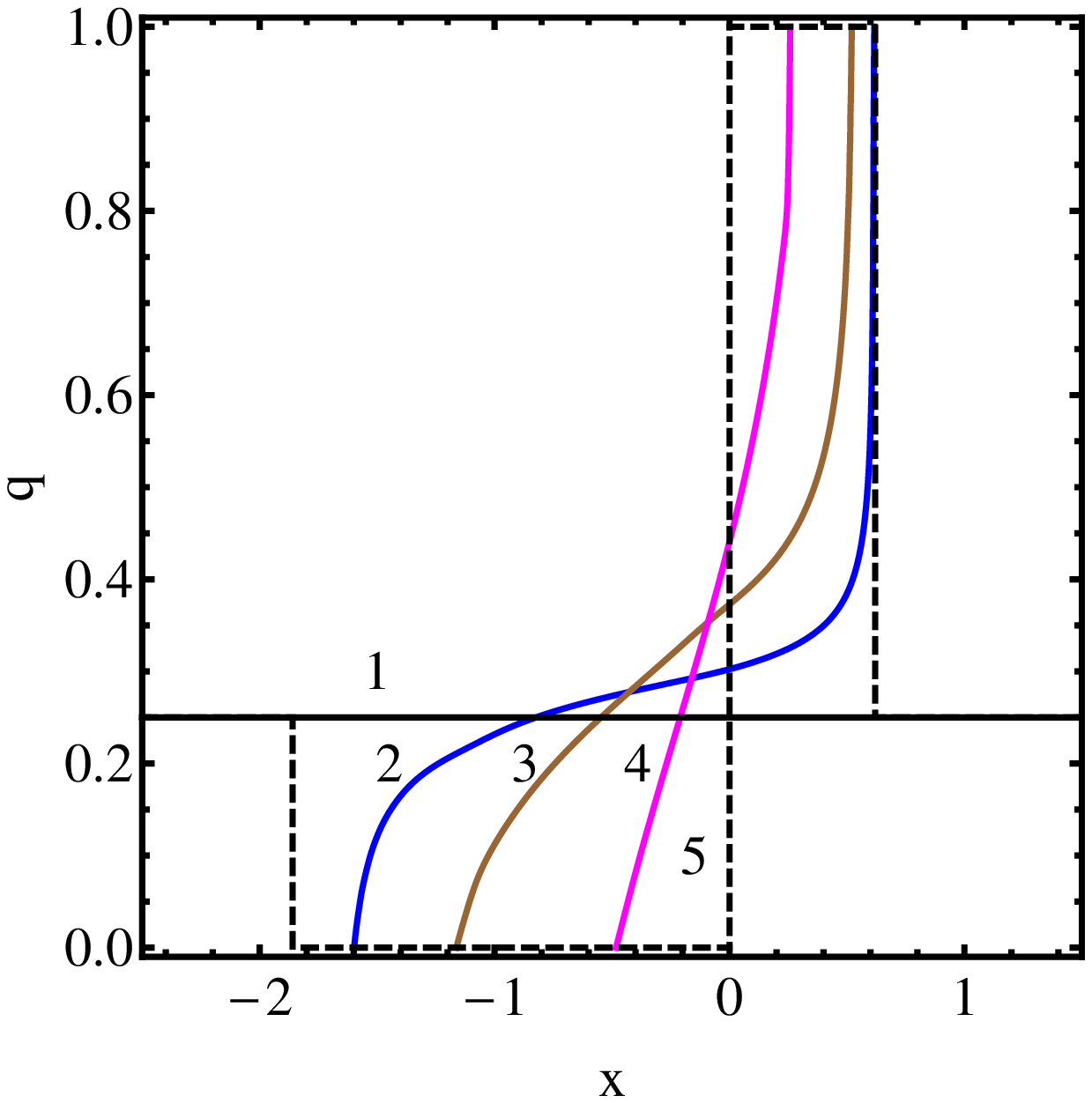}
\includegraphics[scale=0.385] {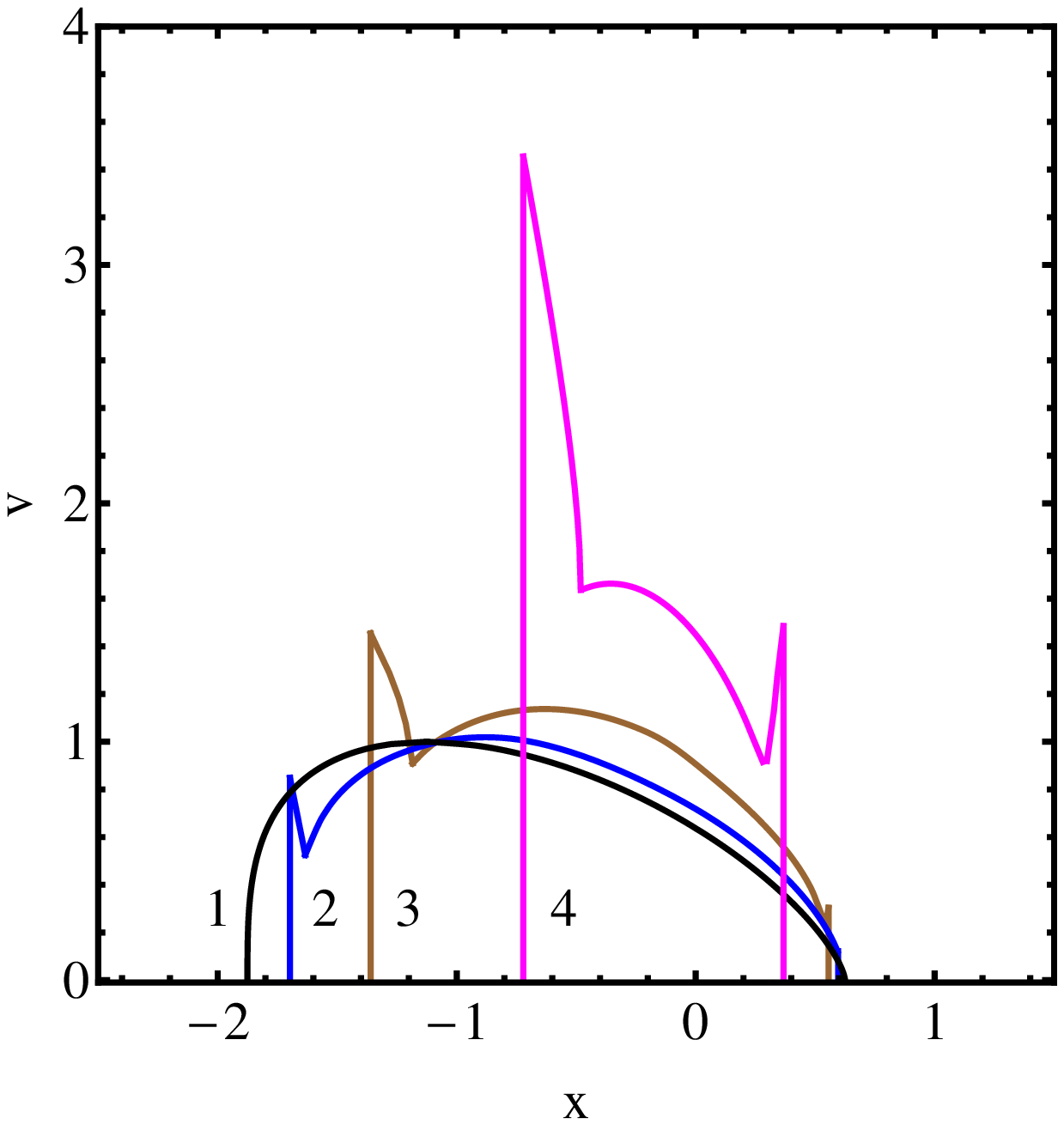}
\includegraphics[scale=0.41] {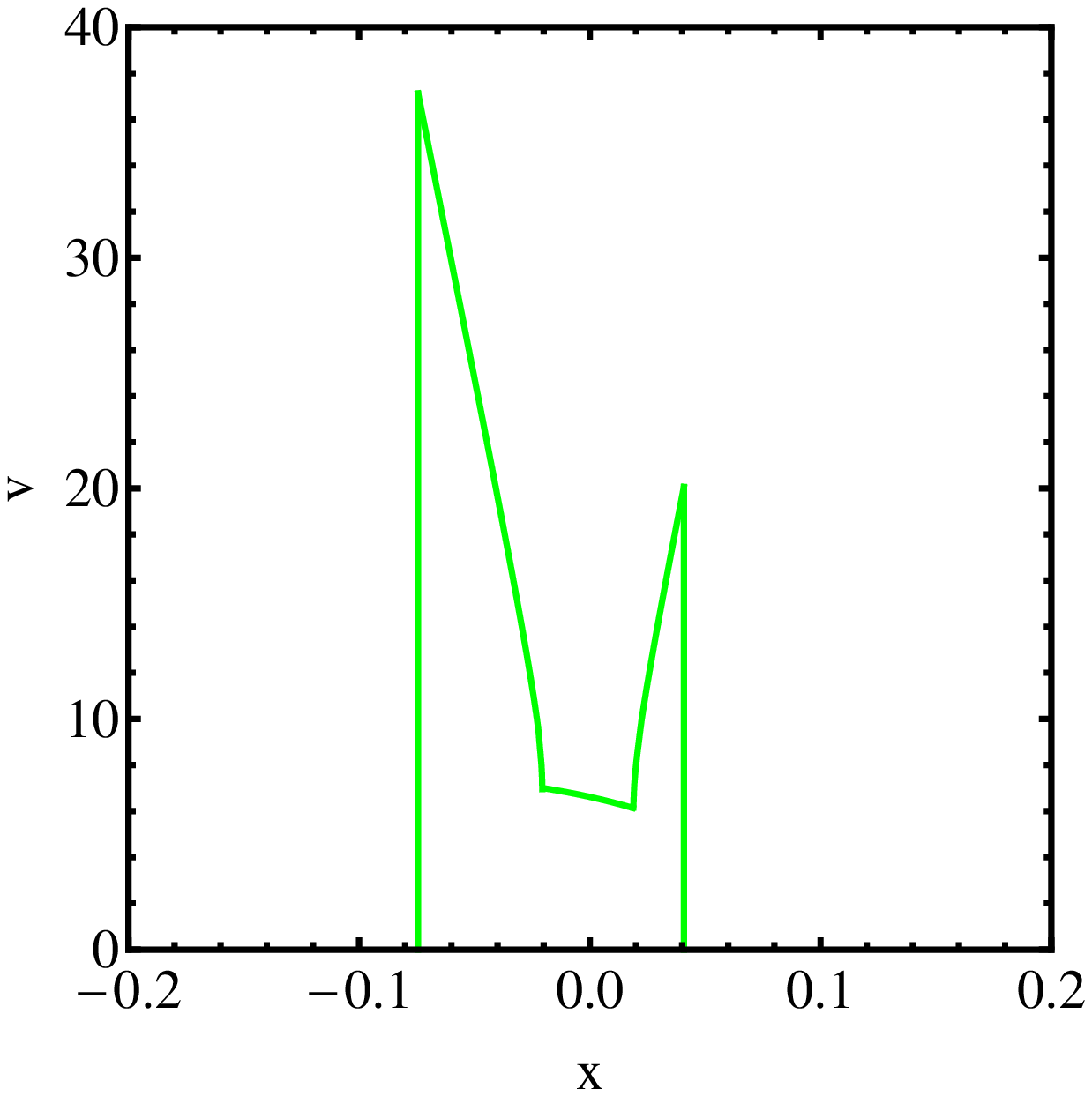}
\caption{The analytical solutions for $q(x,t)$ (left panel) and $v(x,t)$ (middle and right panels) versus $x$  at $n_0=1/4$ and $t=0$ (1), $0.3$ (2), $0.6$ (3), $0.9$ (4), $0.999$ (only the right panel) and $1$ (5) (only the left panel). On the left panel one can see the hodograph region $0<q<1$, the void region $q=0$, the cluster region $q=1$ and the two static regions where $q=q_0$, separated from the hodograph region by standing $q$-shocks. The middle and right panels show $v(x,t)$ in the hodograph region $0<q<1$, in the void region, $q=0$,
and in the cluster region, $q=1$, as well as two traveling $v$-shocks. At $t<0.3$ the cluster region is too small to be seen.
The right panel shows $v(x,t=0.999)$; notice the differences in scales. The parametrization $v_{\text{max}}(x,t=0)=1$ is used.}
\label{qvofx}
\end{figure}

Importantly, only those solutions of Eqs.~(\ref{d1}) and (\ref{d2}) contribute to $s$ where both $q(x,t)[1-q(x,t)]$ and $v(x,t)$ are nonzero, see Eq.~(\ref{action1}), where $\sigma(q)=2q(1-q)$.  There are some special solutions of these equations, however, which are present in the complete solution of the inviscid problem, although they do not contribute to the action in the inviscid limit \cite{MS2014}. These are (i) static solutions $q=\text{const}$ and $v=0$, (ii) a non-trivial void solution: $q=0$ and $v(x,t)\neq 0$,  and (iii) a non-trivial close-packed cluster solution: $q=1$  and $v(x,t)\neq 0$. Actually, there are 4 different static solutions which hold in different regions of space. As a result, the complete inviscid solution includes 7 different regions:

\begin{enumerate}
  \item $q=n_0,\,v=0$ at $x<x_-$.
  \item $q=0,\,v=0$ at $x_-<x<X_{\text{void}}(t)$.
  \item $q=0,\,v\neq 0$ at $X_{\text{void}}(t)<x<x_0(t)$.
  \item $0<q<1,\,v\neq 0$ at $x_0(t)<x<x_1(t)$.
  \item $q=1,\,v\neq 0$ at $x_1(t)<x<X_{\text{cluster}}(t)$.
  \item $q=1,\,v=0$ at $X_{\text{cluster}}(t)<x<x_+$.
  \item $q=n_0,\,v=0$ at $x>x_+$.
\end{enumerate}
These regions can be seen in Fig. \ref{qvofx}. The flow also includes two static shock discontinuities of $q$, located at $x=x_-$ and $x=x_+$, and two moving shock discontinuities of $v$, located at $x=X_{\text{void}}(t)$ and $x=X_{\text{cluster}}(t)$. The quantities $x_-$, $x_+$, $x_0(t)$, $x_1(t)$, $X_{\text{void}}(t)$ and $X_{\text{cluster}}(t)$ will be determined as we go along. All of them depend on $j$ and $n_0$.  Notice that, at t = 0, the solution already includes a pointlike void, $q = 0$, at the point $x =x_-$, and a pointlike close-packed cluster, $q=1$, at $x =x_+$ \cite{MS2014,discontinuities}.

Regions 2 and 3 describe a void, regions 5 and 6 describe a close-packed particle cluster. The dynamics of $v(x,t)$ in regions 3 and 5 is described by two slightly different Hopf equations, following from Eq.~(\ref{d2}) with $\sigma(q)=2q(1-q)$:
\begin{eqnarray}
 \partial_t v + 2v \partial_x v &=& 0\;\;\;\;\text{in region 3}, \label{Hopfvoid}\\
 \partial_t v - 2v \partial_x v &=& 0\;\;\;\;\text{in region 5}. \label{Hopfcluster}
\end{eqnarray}
The solutions of these equations can be written in an implicit form as  \cite{LLfluidmech}
\begin{eqnarray}
 x-2 v t&=& F_{\text{void}}(v)\;\;\;\;\;\;\;\text{in region 3}, \label{Hopfvoidsol}\\
x+2 vt &=& F_{\text{cluster}}(v)\;\;\;\;\text{in region 5}, \label{Hopfclustersol}
\end{eqnarray}
where $F_{\text{void}}(v)$ and $F_{\text{cluster}}(v)$ are functions to be found.  The non-trivial void and cluster solutions (\ref{Hopfvoidsol}) and (\ref{Hopfclustersol}) must be continuously matchable with the hodograph solution at $0<q<1$ which we expose in the following. This constrains the hodograph solution to be regular at $q=0$ and $q=1$.
In addition, this
will enable us to determine the functions $F_{\text{void}}(v)$ and $F_{\text{cluster}}(v)$ once the hodograph solution has been found.

Each of the two static regions $q=n_0,\,v=0$ is separated from the void or cluster regions by a moving shock discontinuity, where $v$ drops from a positive value to zero. These moving shocks are located at a priori unknown points $x=X_{\text{void}}(t)$ and $x=X_{\text{cluster}}(t)$ which satisfy the equations
\begin{eqnarray}
(d/dt) X_{\text{void}}(t) &=& v[X_{\text{void}}(t),t], \label{shockvoid}\\
(d/dt) X_{\text{cluster}}(t) &=& - v[X_{\text{cluster}}(t),t], \label{shockcluster}
\end{eqnarray}
respectively. Here by $v(x,t)$ is meant the void or cluster solution, given  by Eq.~(\ref{Hopfvoidsol}) or (\ref{Hopfclustersol}), respectively. Equations (\ref{shockvoid}) and (\ref{shockcluster}) follow from Eqs.~(\ref{Hopfvoid}) and (\ref{Hopfcluster}) and the conservation
of $v$ \cite{Whitham}.

To reemphasize, neither of the regions described by special solutions contributes to the action, see Eq.~(\ref{action1}) where $\sigma(0)=\sigma(1)=0$.

\section{Hodograph transformation and Laplace's equation}
\label{hodo}

The inviscid  Eqs.~(\ref{d1}) and (\ref{d2}) become linear
upon the  hodograph
transformation, where $q$ and $v$ are treated as the independent variables, and $t$
and $x$ as the dependent ones \cite{LLfluidmech,Courant}. After standard algebra (see Appendix~\ref{hodograph})  Eqs. (\ref{d1}) and (\ref{d2}) become
\begin{eqnarray}
  \partial_v x&=&\sigma^{\prime}(q) v \,\partial_v t-\sigma(q) \partial_q t, \label{Ahod} \\
  \partial_q x&=& -\frac{1}{2} \sigma^{\prime\prime}(q) v^2\, \partial_v t+\sigma^{\prime}(q) v \,\partial_q t. \label{Bhod}
\end{eqnarray}
Differentiating the first equation with respect to $q$, and the second one with respect to $v$, we obtain a linear second-order equation for the function $t(q,v)$:
\begin{equation}\label{eqt}
\sigma (q) \partial_q^2 t-\frac{1}{2} \sigma^{\prime\prime} (q) v^2 \partial_v^2t+2\sigma^{\prime}(q) \partial_q t-2\sigma^{\prime\prime} (q) v \partial_v t=0.
\end{equation}
For the SSEP, $\sigma(q)=2q(1-q)$, Eq.~(\ref{eqt}) becomes
\begin{equation}\label{eqtSSEP}
q(1-q) \partial_q^2 t+v^2 \partial_v^2t+2(1-2q) \partial_q t+4 v \partial_v t=0
\end{equation}
which, for $0<q<1$ and $v\neq 0$, is an elliptic equation \cite{Sommerfeld}. Importantly, this equations admits separation of variables $q$ and $v$.  What are the boundary conditions? First, the value of $v(x,t=0)$ changes, as a function of $x$,
from $0$ to an a priori unknown bounded maximum value $v_0>0$. Employing the invariance of
the inviscid MFT equations under the transformation
$x/\sqrt{\Lambda}\to x$ and $v/\sqrt{\Lambda} \to v$, we can solve the problem for the parametrization
$v_0=1$, calculate the corresponding value of $\Lambda$ in the boundary condition (\ref{delta}),
and then use Eq.~(\ref{fany}) and restore the $\Lambda$-scalings in the final solution. By virtue of the conditions
$q(x,t=0)=n_0$ and $0\leq v (x,t=0)\leq 1$, we demand
\begin{equation}\label{BChod1}
   t= 0 \;\;\;\mbox{at}\;\;\;q=q_0, \;0<v<1.
\end{equation}
An additional boundary condition stems from the fact that $v(x,t=1)$ is a delta-function. As a result,
\begin{equation}\label{BChod2}
t=1 \;\;\;\mbox{at}\;\;\;v\to \infty.
\end{equation}
Finally, $t(q,v)$ must be regular at $v=0$,  $q=0$ and $q=1$, as we observed earlier. The boundary conditions (\ref{BChod1}), (\ref{BChod2}) and the regularity of the solution at $v=0$, $q=0$ and $q=1$
define a Dirichlet problem for the elliptic equation (\ref{eqtSSEP}) and guarantee a unique solution for $t(q,v)$.
Once $t(q,v)$ is found, $x(q,v)$ can be found by integrating any of the relations (\ref{Ahod}) and (\ref{Bhod}).
For example, integrating Eq.~(\ref{Ahod}) over $v$, we obtain
\begin{equation}\label{xintegral}
x(q,v )=
\int_{v}^{\infty}\left[2(2q-1) v \partial_v t +2q(1-q)\partial_q t\right]dv,
\end{equation}
where arbitrary constant is put to zero because $v=\infty$ at $x=0$ (and $t=1$).

\begin{figure}[ht]
\includegraphics[width=2.3in,clip=]{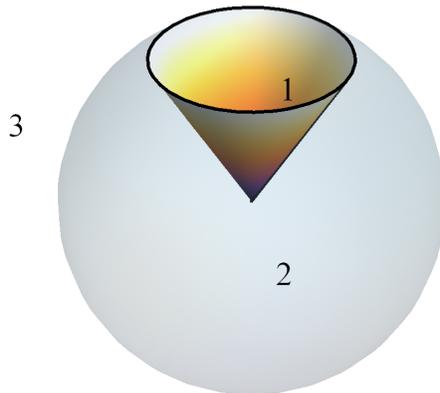}
\caption{Geometry of the Dirichlet problem in the extended hodograph space. The solution has different forms in
regions 1, 2 and 3.}
\label{cone}
\end{figure}

Remarkably, a simple change of variables reduces Eq.~(\ref{eqtSSEP}) to the Laplace's equation in an extended (three-dimensional) space \cite{Trubnikov}. Indeed, let us introduce
a new independent variable $\theta=\arccos (1-2q)$. As $0\leq q\leq 1$, we have $0\leq \theta \leq \pi$.  Equation (\ref{eqtSSEP}) becomes
\begin{equation}\label{eqtSSEPnew}
   \partial_{\theta}^2 t +3 \cot \theta \,\partial_{\theta} t +r^2 \partial_r^2 t +4 r \,\partial_r t=0,
\end{equation}
where we have renamed $v$ by $r$. Finally, we introduce an auxiliary angle $\phi$, so that $0\leq \phi \leq 2\pi$, and define a new dependent variable
\begin{equation}\label{Psi}
\Psi(r,\theta,\phi)=r \,[t(r,\theta)-1] \sin \theta \cos \phi.
\end{equation}
As one can check, using Eq.~(\ref{eqtSSEPnew}), the function $\Psi(r,\theta,\phi)$
obeys the Laplace's equation,
\begin{equation}\label{3DLaplace}
    \nabla^2 \Psi=0
\end{equation}
in the spherical coordinates $r,\theta$ and $\phi$. This opens the way to a full analytical solution of the
inviscid problem. The boundary conditions for the harmonic function  $\Psi(r,\theta,\phi)$
stem from the boundary conditions for $t(q,v)$. The boundary condition (\ref{BChod1}) becomes
\begin{equation}\label{Psi1}
    \Psi=-r \sin \theta_{0} \cos \phi\;\;\;\;\text{on the conical surface}\;\;\;\;\theta=\theta_{0},\;\;  0\leq r\leq 1,
\end{equation}
where $\theta_{0}=\arccos (1-2 n_0)$.  As we will see, $t$ approaches $1$ sufficiently
rapidly. Therefore, the boundary condition (\ref{BChod2}) becomes
\begin{equation}\label{infinity}
    \Psi\to 0 \;\;\;\text{at}\;\;\; r \to \infty,
\end{equation}
in spite of the presence of $r$-factor in Eq.~(\ref{Psi}).  Finally, the regularity of $t(q,v)$ at $q=0$ and $q=1$ yields the condition
\begin{equation}\label{regular}
    |\Psi(r,\theta,\phi)|<\infty\;\;\;\text{at}\;\;\;\theta=0\;\;\; \text{and} \;\;\; \theta=\pi.
\end{equation}
The Dirichlet problem, defined by Eqs.~(\ref{3DLaplace})-(\ref{regular}), has a unique solution we are going to find.  In view of Eq.~(\ref{Psi1}) the $\phi$-dependence of $\Psi(r,\theta,\phi)$ is simply
$\cos \phi$. Figure \ref{cone} shows the geometry of the Dirichlet problem. Note that, because of the intrinsic particle-hole symmetry of the SSEP process, we can only consider the case of $n_0\leq 1/2$, where $0\leq \theta_0\leq \pi/2$. For $n_0=1/2$ one has $\theta_{0}=\pi/2$, and the conical surface $\theta=\theta_{0},\;\;  0\leq r\leq 1$ degenerates into a disk. In this special case the solution of the Dirichlet problem can be obtained in elementary functions by employing elliptic coordinates \cite{MS2014}.

\section{Hodograph solution}
\label{solution}
\subsection{Solving the Dirichlet problem}
\label{spherical}

We will interpret $\Psi$ as electric potential and expand it in the proper eigenfunctions of the Laplacian operator in spherical coordinates $r, \theta$ and $\phi$.  For the geometry in question (see Fig.~\ref{cone})
we need to use three different expansions: in the region $0<r<1,\,0\leq \theta < \theta_{0}$ (inside the cone:
region 1), in the region $0<r<1,\,\theta_{0} < \theta \leq \pi$ (outside the cone, but inside the unit sphere: region 2), and in the region $1<r<\infty,\,0\leq \theta \leq \pi$ (outside the unit sphere: region 3). The expansions are
\begin{eqnarray}
  \Psi_1(r,\theta,\phi) &=& -r \sin \theta \cos \phi+\cos \phi \sum_{k=1}^{\infty} A_k r^{\alpha_k} P_{\alpha_k}^1 (\cos \theta),  \label{region1}\\
\Psi_2(r,\theta,\phi) &=& -r \sin \theta \cos \phi+\cos \phi \sum_{k=1}^{\infty} B_k r^{\beta_k} P_{\beta_k}^1 (-\cos \theta),  \label{region2}\\
\Psi_3(r,\theta,\phi)&=& \cos \phi \sum_{l=1}^{\infty} C_l r^{-l-1} P_l^1 (\cos \theta),  \label{region3}
\end{eqnarray}
where the subscripts $1,2$ and $3$ denote the corresponding regions, and $P_{\nu}^1(\cos \theta)$ is the associated Legendre function of the first kind \cite{Abramowitz}. The first term on the right hand side of each of the equations~(\ref{region1}) and (\ref{region2}) is in itself a harmonic function obeying the inhomogeneous boundary condition (\ref{Psi1}). Therefore, each of the terms
$k=1,2,\dots$ in the sums over $k$ in Eqs.~(\ref{region1}) and (\ref{region2}) must vanish at $\theta=\theta_{0}$:
\begin{equation}\label{eigenvalues}
 P_{\alpha_k}^1 (\cos \theta_0) =0 \;\;\;\text{and}\;\;\; P_{\beta_k}^1 (-\cos \theta_0) =0,\;\;\;k=1,2,\dots
\end{equation}
These conditions determine discrete spectra $\alpha_k$ and $\beta_k$. For example, for $n_0=1/4$ one has $\theta_{0}=\pi/3$, and
$\alpha_1=3.1956\dots$, $\alpha_2=6.2195\dots$, $\alpha_3=9.2288\dots$, etc, whereas $\beta_1=1.4241\dots$, $\beta_2=2.9043\dots$, $\beta_3=4.3957\dots$, etc. The multipole expansion $\Psi_3$ from Eq.~(\ref{region3}) obeys the boundary condition (\ref{infinity}). As one can see, the leading far-field multipole is a dipole,  $\Psi_3\sim r^{-2}$.

To complete the solution in terms of expansions~(\ref{region1})-(\ref{region3}), we must find
the coefficients $A_k$, $B_k$ and $C_k$.
We will do it by first determining the effective charge distribution on the (non-conducting) conical surface $\theta=\theta_{0},\;\;  0\leq r\leq 1.$ Let the a priori unknown volume density of this charge distribution be $w(\mathbf{r})$. Then the potential $\Psi$ can be written as
\begin{equation}\label{Poisson1}
    \Psi(\mathbf{r}) = -\int \,\frac{w(\mathbf{r}^{\prime}) d^3\mathbf{r}^{\prime}}{|\mathbf{r}-\mathbf{r}^{\prime}|}.
\end{equation}
Now, $w(\mathbf{r})$ can be sought as
\begin{equation}\label{bulkchargedensity}
w(r,\theta,\phi)= - \frac{1}{r} \,\chi (r,\theta_{0}) \cos \phi \,\delta(\theta-\theta_{0}), \;\;\;0\leq r \leq 1.
\end{equation}
The corresponding surface charge density on the conical surface is $-\chi(r,\theta_{0}) \cos \phi$, with an a priori unknown $\chi(r,\theta_{0})>0$. Plugging Eq.~(\ref{bulkchargedensity}) into Eq.~(\ref{Poisson1}) and performing the  integration over $\theta^{\prime}$, we obtain
\begin{widetext}
\begin{eqnarray}
  \Psi(r,\theta,\phi) &=& -\sin \theta_{0} \int_0^1 dr^{\prime} r^{\prime} \chi (r^{\prime},\theta_{0}) \int_0^{2\pi} \frac{d\phi^{\prime}\,\cos \phi^{\prime}}{\{r^2+r^{\prime 2}-2 r r^{\prime} \left[ \cos \theta \cos \theta_{0}+\sin \theta \sin \theta_{0} \cos (\phi-\phi^{\prime})\right]\}^{1/2}} \nonumber \\
  &=& -\cos \phi \sin \theta_{0} \int_{0}^{1} dr^{\prime} r^{\prime} \chi(r^{\prime},\theta_{0})\int_{0}^{2\pi}  \frac{d\xi\,\cos\xi}{\sqrt{r^{2}+r^{\prime 2}-2rr^{\prime}\cos\Gamma(\theta,\xi,\theta_{0})}},
  \label{Poissonspherical}
\end{eqnarray}
\end{widetext}
where $\cos\Gamma(\theta,\xi,\theta_{0})=\cos\theta\cos\theta_{0}+\sin\theta\sin\theta_{0}\cos\xi$.

To determine $\chi(r,\theta_0)$ in terms of the unknown coefficients  $A_k$ and $B_k$, we consider the regions 1 and 2 and apply the Gauss' law to an infinitesimally small volume which includes an infinitesimally  small element of the conical surface:
\begin{equation}\label{Gauss10}
   \left(\frac{1}{r}\frac{\partial \Psi_2}{\partial \theta} -\frac{1}{r}\frac{\partial \Psi_1}{\partial \theta} \right)\Big|_{\theta=\theta_0} =4 \pi\chi(r,\theta_0) \cos \phi,
\end{equation}
where $\Psi_1$ and $\Psi_2$ are given by Eqs. (\ref{region1}) and (\ref{region2}), respectively. Performing the differentiation on the left side of Eq.~(\ref{Gauss10}), we use the relation (see Ref. \cite{Abramowitz})
\begin{equation*}
\frac{d}{dx}P^{1}_{\nu}(x)=-\frac{\nu}{1-x^{2}}P^{1}_{\nu+1}(x)+\frac{\nu+1}{1-x^{2}}xP^{1}_{\nu}(x).
\end{equation*}
As $P^{1}_{\alpha}(\cos\theta_0)=0$ and $P^{1}_{\beta}(-\cos\theta_0)=0$, we obtain, for $\theta=\theta_{0}$:
\begin{equation*}
\partial_{\theta}P^{1}_{\alpha_{k}}(\cos\theta)=\frac{\alpha_{k}}{\sin\theta_{0}}P^{1}_{\alpha_{k}+1}(\cos\theta_{0}),\quad \partial_{\theta}P^{1}_{\beta_{k}}(-\cos\theta)=-\frac{\beta_{k}}{\sin\theta_{0}}P^{1}_{\beta_{k}+1}(-\cos\theta_{0}),
\end{equation*}
and so
\begin{eqnarray*}
  \frac{1}{r}\frac{\partial \Psi_2}{\partial \theta}\Big|_{\theta=\theta_0} &=& -\cos\theta_0\cos\phi-\frac{\cos\phi}{\sin\theta_{0}}
\sum_{k=1}^{\infty}B_{k}\beta_{k}r^{\beta_{k}-1}P^{1}_{\beta_{k}+1}(-\cos\theta_{0}), \\
  \frac{1}{r}\frac{\partial \Psi_1}{\partial \theta} \Big|_{\theta=\theta_0}&=& -\cos\theta_0\cos\phi+\frac{\cos\phi}{\sin\theta_{0}}
\sum_{k=1}^{\infty}A_{k}\alpha_{k}r^{\alpha_{k}-1}P^{1}_{\alpha_{k}+1}(\cos\theta_{0}).
\end{eqnarray*}
Now Eq.~(\ref{Gauss10}) becomes
\begin{equation*}
-4\pi\sin \theta_{0} \,\chi(r,\theta_0)=\sum_{k=1}^{\infty}A_{k}\alpha_{k}r^{\alpha_{k}-1}P^{1}_{\alpha_{k}+1}(\cos\theta_{0})+
\sum_{k=1}^{\infty}B_{k}\beta_{k}r^{\beta_{k}-1}P^{1}_{\beta_{k}+1}(-\cos\theta_{0}),
\end{equation*}
and we obtain
\begin{equation}
\label{sigma}
\chi(r,\theta_{0})=\sum_{k=1}^{\infty} \left(a_{k}r^{\alpha_{k}-1}+b_{k}r^{\beta_{k}-1}\right)
=\sum_{k=1}^{\infty}c_{k}r^{\gamma_{k}-1},\;\;\;\;0\leq r \leq 1.
\end{equation}
Here we have denoted for brevity
\begin{eqnarray}
 a_{k} &=& -\frac{\alpha_{k}P^{1}_{\alpha_{k}+1}(\cos\theta_{0})}{4\pi\sin\theta_{0}}A_{k}, \label{coeffA} \\
   b_{k} &=& -\frac{\beta_{k}P^{1}_{\beta_{k}+1}(-\cos\theta_{0})}{4\pi\sin\theta_{0}}B_{k}. \label{coeffB}
\end{eqnarray}
Further, $\gamma_k=\gamma_k(\theta_{0})\geq 1$ in Eq.~(\ref{sigma}) is the union of eigenvalues $\alpha_{k}$ and $\beta_{k}$ ($\gamma_k$ can be ordered so that $\gamma_k$ grows monotonically with $k$)  \cite{union}, while $c_k$ is the corresponding union of $a_k$ and $b_k$.  Plugging Eq.~(\ref{sigma}) into Eq.~(\ref{Poissonspherical}) and changing the order of integration
over $r^{\prime}$ and $\xi$, we obtain
\begin{equation}\label{10}
\Psi(r,\theta,\phi)=-\cos\phi\sin\theta_{0}\int_{0}^{2\pi}d\xi\,\cos\xi \, \sum_{k=1}^{\infty}c_{k} \,\int_{0}^{1}\frac{dr^{\prime}r^{\prime \gamma_{k}}}{\sqrt{r^{2}+r^{\prime 2}-2rr^{\prime}\cos\Gamma(\theta_{0},\theta,\xi)}}.
\end{equation}
Evaluating this expression on the conical surface and using the
boundary condition~(\ref{Psi1}), we arrive at an infinite set of linear algebraic equations for
the coefficients $c_k=c_k(n_0)$ which can be solved analytically. The solution, presented in
Appendix~\ref{sigmacoeff}, yields $c_k$ in terms of an infinite product of factors including the
eigenvalues $\gamma_k$, see  Eq.~(\ref{c_n solution}). Having found $c_k$, we return to $A_k$ and $B_k$,
\begin{equation}\label{coeffABsecond}
A_{k}=-\frac{8\pi\sqrt{n_0(1-n_0)} \,a_{k}}{\alpha_{k}P^{1}_{\alpha_{k}+1}(1-2n_0)},\;\;\;\;\;
B_{k}=-\frac{8\pi\sqrt{n_0(1-n_0)}\,b_{k}}{\beta_{k}P^{1}_{\beta_{k}+1}(2n_0-1)},
\end{equation}
and use Eq.~(\ref{10}) to determine $C_k$ entering Eq.~(\ref{region3}). The latter calculation is straightforward. Indeed, the
expression
\begin{equation*}
\frac{1}{\sqrt{r^{2}+r^{\prime 2}-2rr^{\prime}\cos\Gamma(\theta_{0},\theta,\xi)}}
\end{equation*}
is the generating function of Legendre polynomials \cite{Jackson}. Expanding it in the Legendre polynomials in region 3, we can
evaluate the double integral in Eq.~(\ref{10}), see Appendix~\ref{Psicoeff}1. The result is
\begin{equation}\label{C_lsecond}
C_{l}=-\frac{4\pi\sqrt{n_0(1-n_0)}\,P^{1}_{l}(1-2n_0)}{l(l+1)}\sum_{k=1}^{\infty}\frac{c_{k}}{\gamma_{k}+l+1}.
\end{equation}
Equations~(\ref{region3}), (\ref{C_lsecond}) and (\ref{c_n solution}) completely determine $\Psi_3(r,\theta,\phi)$ from Eq.~(\ref{region3}). In their turn, $\Psi_1(r,\theta,\phi)$ and $\Psi_2(r,\theta,\phi)$ are completely determined by Eqs. (\ref{region1}), (\ref{region2}),  (\ref{akm}), (\ref{bkm}) and (\ref{coeffABsecond}).

\subsection{Calculating $t(q,v)$ and $x(q,v)$}
\label{hodsol}
Now we can calculate the hodograph solutions $t(q,v)$ and $x(q,v)$.
Consider first $v \geq 1$ which corresponds to region 3.  Here Eqs. (\ref{Psi}) and (\ref{region3}) yield
\begin{equation}\label{tmore}
t(q,v \geq 1)=1+\sum_{l=1}^{\infty}\,\frac{C_{l} v^{-l-2}P^{1}_{l}(1-2q)}{2\sqrt{q(1-q)}},
\end{equation}
where we have substituted $\sin\theta=2\sqrt{q(1-q)}$ and returned from $r$ to $v$.   Calculating the partial derivatives $\partial_q t$ and $\partial_v t$ and using Eq.~(\ref{xintegral}), we obtain
after some algebra (see Appendix~\ref{xhodmore}):
\begin{equation}\label{xmore}
x(q,v \geq 1)=\frac{1}{2\sqrt{q(1-q)}}
\sum_{l=1}^{\infty}\frac{lP^{1}_{l+1}(1-2q)+(1-2q)(l+2)P_{l}^{1}(1-2q)}{l+1}C_{l}v^{-l-1}.
\end{equation}

\begin{figure}
\includegraphics[scale=0.9]{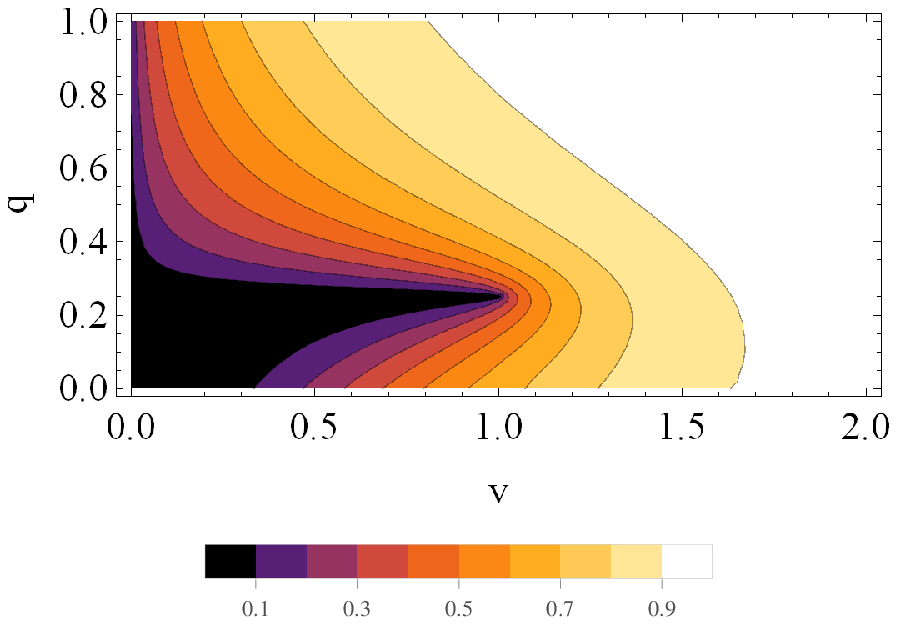}
\includegraphics[scale=0.9]{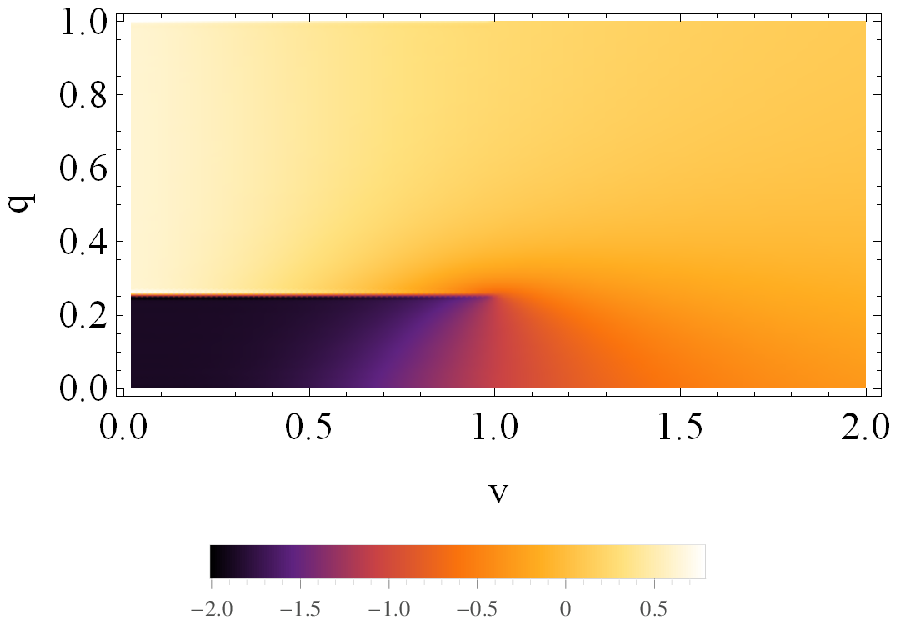}
\caption{A contour-density plot of $t(q,v)$ (left panel) and a density plot of $x(q,v)$ (right panel) in the hodograph region
$0 \leq q \leq 1$ for $n_0=1/4$, that is $\theta_{0}=\pi/3$. The parametrization $v_{\text{max}}(x,t=0)=1$
is used.}
\label{txcountrplt}
\end{figure}

The region $0\leq v\leq1$ includes two sub-regions, $0\leq q \leq n_0$ and $n_0\leq q\leq 1$, corresponding to regions 1 and 2 of the Dirichlet problem, respectively. Employing Eqs. (\ref{Psi}), (\ref{region1}) and (\ref{region2}), we obtain
\begin{eqnarray}
  t(0\leq q \leq n_0,0 \leq v\leq 1) &=& \sum_{k=1}^{\infty}\frac{A_{k}v^{\alpha_{k}-1}P^{1}_{\alpha_{k}}(1-2q)}{2\sqrt{q(1-q)}}, \label{tlessl}\\
 t(n_0 \leq q \leq 1,0 \leq v \leq 1)&=& \sum_{k=1}^{\infty}\frac{B_{k}v^{\beta_{k}-1}P^{1}_{\beta_{k}}(2q-1)}{2\sqrt{q(1-q)}}. \label{tlessm}
\end{eqnarray}
To obtain $x(q,v)$, we calculate the partial derivatives $\partial_q t$ and $\partial_v t$ in the two subregions,
plug them into Eq. (\ref{xintegral}) and perform the integrations over $v$, see Appendix~\ref{xhodless}.
The result is
\begin{eqnarray}
x(0 \leq q \leq n_0, 0 \leq v \leq 1)&=&\frac{1}{2\sqrt{q(1-q)}}\sum_{k=1}^{\infty}\Big\{A_{k}(v^{\alpha_{k}}-1)
[3(1-2q)P^{1}_{\alpha_{k}}(1-2q)-P^{1}_{\alpha_{k}+1}(1-2q)] \Big. \nonumber \\
&+&\left.\frac{kP^{1}_{k+1}(1-2q)+(1-2q)(k+2)P_{k}^{1}(1-2q)}{1+k}C_{k}\right\}, \label{xlessl} \\
x(n_0 \leq q \leq 1, 0 \leq v \leq 1)&=&\frac{1}{2\sqrt{q(1-q)}}\sum_{k=1}^{\infty}\Big\{B_{k}(v^{\beta_{k}}-1)
[3(1-2q)P^{1}_{\beta_{k}}(2q-1)+P^{1}_{\beta_{k}+1}(2q-1)] \nonumber \\
&+&\frac{kP^{1}_{k+1}(1-2q)+(1-2q)(k+2)P_{k}^{1}(1-2q)}{1+k}C_{k}\Big\}, \label{xmr}
\end{eqnarray}
Equations~(\ref{tmore})-(\ref{xmr}) completely describe the hodograph solution for arbitrary $0<n_0<1$ and
$v_0\equiv \max v(x,0) =1$.
Figure \ref{txcountrplt} shows two-dimensional plots of  $t(q,v)$ and  $x(q,v)$ for $n_0=1/4$ which corresponds to $\theta_{0}=\pi/3$. Inverting $t(q,v)$ and $x(q,v)$ (this can only be done numerically), one obtains $q(x,t)$ and $v(x,t)$ for this solution. Figure~\ref{qvofx} shows plots of $q$ and $v$ versus $x$ at different times for  $n_0=1/4$.

\subsection{Calculating $s_*$, $j_*$ and $\Lambda_*$}
\label{ljs}

Now we can determine the action $s_*$,  the integrated current $j_*$ and the Lagrange multiplier $\Lambda_*$ for our parametrization $v_0 \equiv \max v(x,0) =1$.

\subsubsection{Calculating the action $s_*$}
The action $s_*$ can be found from Eqs.~(\ref{Ham}) and (\ref{action1}). It is convenient to evaluate
$s_*=H_0$ at $t\to 1$. Here we can account only for the leading and subleading terms in Eq.~(\ref{tmore}):
\begin{equation}\label{ttto1}
t(q,v \gg 1)=
1-\frac{C_1}{v^3}.
\end{equation}
(The subleading term corresponds to the dipole asymptotic of the potential $\Psi_3$.)
In its turn, the leading term of Eq.~(\ref{xmore}) at $t\to 1$ is
\begin{equation}\label{xtto1}
x(q,v\gg 1)\simeq \frac{3C_1(2q-1)}{v^2}.
\end{equation}
Equations~(\ref{ttto1}) and (\ref{xtto1}) yield self-similar asymptotics of $v(x,t)$ and $q(x,t)$, following from the hodograph
solution at $t\to 1$:
\begin{equation}\label{similarityv}
v(x,t\to 1)=\frac{C_1^{1/3}}{(1-t)^{1/3}},\;\;\;|x|\leq 3\ell(t),
\end{equation}
and
\begin{equation}\label{similarityq}
 q(x,t\to 1)=\frac{1}{2}\left[1+\frac{x}{3\ell(t)}\right],\;\;\;|x|\leq 3 \ell(t),
\end{equation}
where $\ell(t)=C_1^{1/3}(1-t)^{2/3}$ is the dynamic length scale of the solution at $t\to 1$.  In compliance with the boundary condition (\ref{delta}),
$v(x,t)=\partial_x p(x,t)$ blows up at $x=0$ at $t=1$.  As one can see from Eqs.~(\ref{similarityv}) and (\ref{similarityq}), the flow becomes symmetric as $t\to 1$. More precisely, $v$ develops a plateau, whereas $q-1/2$ is an odd function of $x$.  As one can check, Eqs.~(\ref{similarityv}) and (\ref{similarityq}) solve the inviscid MFT equations (\ref{d1}) and (\ref{d2}) exactly.
Now we can calculate
\begin{equation}\label{s*}
s_*=\int_{-3\ell(t)}^{3\ell(t)} dx\,q(x,t\to 1) [1-q(x,t\to 1)] v^2(x,t\to 1) =C_1 =4 \pi n_0 (1-n_0)\sum_{k=1}^{\infty} \frac{c_k}{\gamma_k+2},
\end{equation}
where we have used Eq.~(\ref{C_lsecond}). To remind the reader,  $c_k$ is given
by Eq.~(\ref{c_n solution}).

\subsubsection{Calculating the integrated current $j_*$}
At $t=0$ the inviscid solution already includes a point-like void where $q=0$, and a point-like cluster where $q=1$ \cite{MS2014}. The positions of the point-like void and cluster, $x_-$ and $x_+$, coincide with the points
where $v(x,t=0)=0$. We can find these points from  Eqs.~(\ref{xlessl}) and (\ref{xmr}), respectively, by evaluating them at $q=n_0$ and $v=0$. After some algebra, we find
\begin{equation}\label{x1}
x_{-}=-4\pi(1-n_0)\sum_{k=1}^{\infty}\frac{c_{k}}{\gamma_{k}},\;\;\;\;\;\;
x_{+}=4\pi n_0\sum_{k=1}^{\infty}\frac{c_{k}}{\gamma_{k}}.
\end{equation}
The integrated current $j_*$ is equal to
\begin{equation}\label{j*}
    j_* = n_0 |x_-| =  4 \pi n_0 (1-n_0) \sum_{k=1}^{\infty}\frac{c_k}{\gamma_k}.
\end{equation}

\subsubsection{Calculating the Lagrange multiplier $\Lambda_*$}

To calculate $\Lambda_*$, we can employ the conservation law $\int_{-\infty}^{\infty} v(x,t) \,dx = \Lambda_* = \text{const}$. The integral $\int_{-\infty}^{\infty} v(x,t) \,dx$ can be conveniently calculated at $t=0$, where $v(x,0)$ is fully described by the hodograph solution, Eqs.~(\ref{xlessl}) and (\ref{xmr}), where we set $q=n_0$. Note that the function $v(x,0)$ is single valued and has a single maximum (equal to 1) at some point $x=x_m$ which depends on $n_0$. Its inverse function $x(v)$, however, has two branches. We denote them as $x_{<}(v)$ for $x_-<x<x_m$, and $x_{>}(v)$ for $x_m <x<x_+$. Instead of integrating $v(x,0)$ over $x$, we can integrate $x_>(v)-x_<(v)$  over $v$:
\begin{equation}\label{Lambda*}
    \Lambda_*=\int_{x_-}^{x_+} v(x,0) dx=\int_{0}^{1}\left[x_{>}(v)-x_{<}(v)\right] dv.
\end{equation}
Using Eqs.~(\ref{xlessl}) and (\ref{xmr}) for $x_<(v)$ and $x_<(v)$, respectively, we finally obtain
\begin{equation}\label{Lambda*1}
\Lambda_*=4\pi \sum_{k=1}^{\infty}\frac{c_{k}}{\gamma_{k}+1}.
\end{equation}

\subsection{Calculating $\Phi(n_0)$}

\begin{figure}
\includegraphics[scale=0.7] {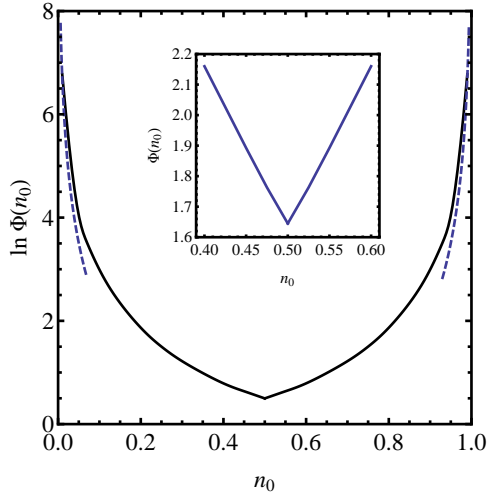}
\caption{The natural logarithm of the function $\Phi(n_0)$ versus $n_0$. Dashed curves: the $n_0\to 0$ and $n_0 \to 1$ asymptotics (\ref{PhiRW}) and (\ref{PhiRWholes}), corresponding to independent random walk of particles and holes, respectively.  Inset: a blowup of the density region close to the half-filling density $n_0=1/2$, showing
the singularity $\Phi(n_0)\sim |n_0-1/2|$.}
\label{Phifig}
\end{figure}

Using Eqs.~(\ref{fany}),(\ref{s*}), (\ref{j*}) and (\ref{c_n solution}), we obtain
\begin{equation}\label{Phi}
    f(n_0,n_0)\equiv \Phi(n_0) = \frac{\tilde{s}}{\tilde{j}^3},
\end{equation}
where
\begin{eqnarray}
\tilde{s} &=&  \left(\gamma_1-1\right)\sum_{k=1}^{\infty} \frac{(\gamma_{k}-k-1)(\gamma_{k+1}-1)}{k(\gamma_{k}+2)} \prod_{l=1}^{\infty}\!\!\!\!~^{~^\prime}   \frac{(\gamma_{k}-l-1)(\gamma_{l+1}-1)}{l(\gamma_k-\gamma_l)} , \label{tildes}\\
 \tilde{j} &=& \left(\gamma_1-1\right)\sum_{k=1}^{\infty} \frac{(\gamma_{k}-k-1)(\gamma_{k+1}-1)}{k\gamma_k} \prod_{l=1}^{\infty}\!\!\!\!~^{~^\prime}
    \frac{(\gamma_{k}-l-1)(\gamma_{l+1}-1)}{l(\gamma_k-\gamma_l)} , \label{tildej}
\end{eqnarray}
where the symbol ``$~^\prime~$'' means that the multiplier with $k=l$ is skipped. Equations~(\ref{Phi})-(\ref{tildej}), alongside with the relation
\begin{equation}
\ln {\cal P}(J,T,n_0) \simeq -\frac{\Phi(n_0) J^3}{T},\;\;\;\;\frac{J}{\sqrt{T}} \gg 1,
\label{result}
\end{equation}
is a central result of this work.  As one can see, $\Phi(n_0)$ only depends on the eigenvalues $\gamma_1 (n_0)$, $\gamma_2 (n_0)$, $\dots$. Figure~\ref{Phifig}
shows the dependence of $\Phi$ on $n_0$. The value at half-filling, $\Phi(n_0=1/2)=\pi^2/6$ was previously calculated in Ref.~\cite{MS2014}, see also Ref. \cite{DG2009b}. As expected, $\Phi(n_0)$ is symmetric with respect to the half-filling density. It diverges at $n_0\to 0$ and $n_0 \to 1$ as expected. Indeed, at $n_0 \ll 1$ the inviscid solution approaches that for the non-interacting random walkers, where
\begin{equation}\label{PhiRW}
\Phi(n_0 \ll 1)=\frac{1}{12 n_0^2},
\end{equation}
see Refs. \cite{DG2009b} and \cite{MS2014} and Eq.~(\ref{PRW}). Similarly, at $1-n_0 \ll 1$ the \emph{holes} behave as non-interacting random walkers, and we obtain
\begin{equation}\label{PhiRWholes}
\Phi(1-n_0\ll 1)=\frac{1}{12 (1-n_0)^2}.
\end{equation}
The asymptotics (\ref{PhiRW}) and (\ref{PhiRWholes}) are shown as dashed lines in Fig.~\ref{Phifig}.
One can also see that $\Phi(n_0)$ exhibits a singularity $\Phi \sim |n_0-1/2|$ at the half-filling density $n_0=1/2$, see Fig. \ref{Phifig} and its inset. We argue that this singularity  only appears in the inviscid limit. It should be smoothed out by diffusion in the exact large deviation function  $s(j,n_0,n_0)$.

\section{Hopf solutions}
\label{solhopf}

Although the void and cluster regions do not contribute to the action, they are important attributes of the optimal path of the system, so we will present them now.  The dynamics of $v(x,t)$ in the void region, $q=0$, and in the cluster region, $q=1$, is determined by the solutions
Eqs.~ (\ref{Hopfvoidsol}) and (\ref{Hopfclustersol}) of the Hopf equations (\ref{Hopfvoid}) and (\ref{Hopfcluster}), respectively. To find the functions
$F_{\text{void}}(v)$ and $F_{\text{cluster}}(v)$, we will continuously match these solutions with the hodograph solution at $q=0$ and $q=1$.

\subsection{Void region}
\begin{figure}[ht]
\includegraphics[scale=0.5]{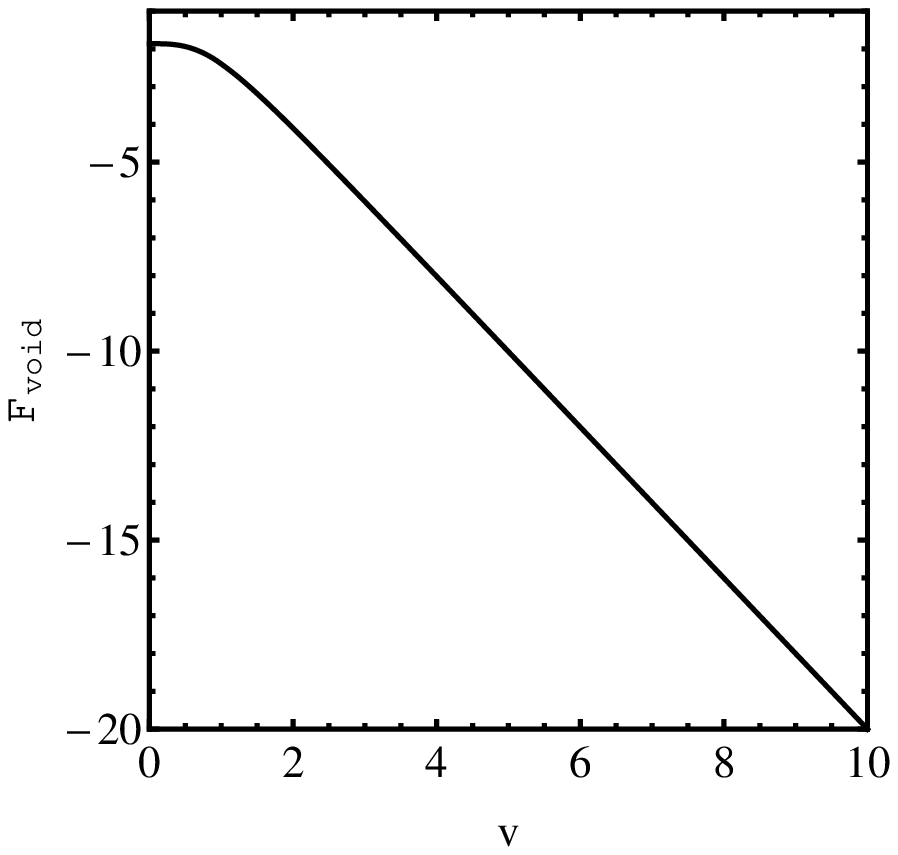}
\includegraphics[scale=0.48]{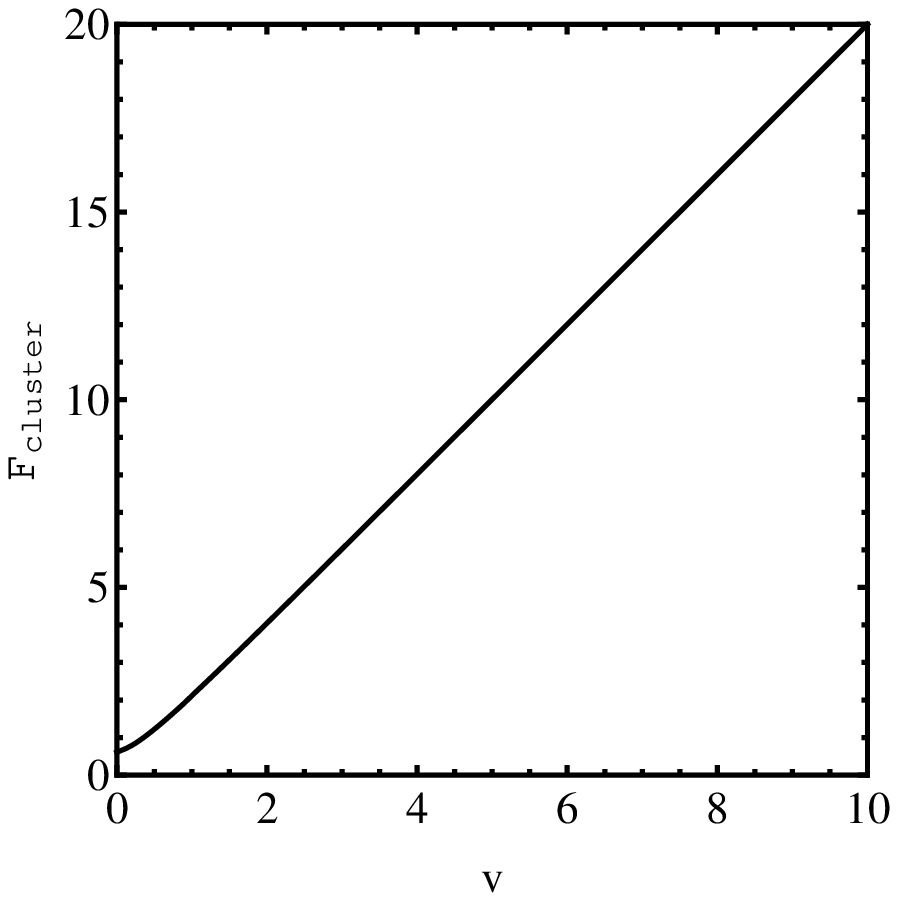}
\caption{The functions $F_{\text{void}}(v)$ (left panel) and $F_{\text{cluster}}(v)$ (right panel) for $n_0=1/4$. The parametrization $v_{\text{max}}(x,t=0)=1$ is used.}
\label{FcustrFvoid}
\end{figure}

Let us evaluate the hodograph solution Eq. (\ref{tlessl}) for $t(q,0\leq v\leq 1)$ at $q=0$:
\begin{equation}\label{tq0l}
t(q=0,0 \leq v \leq 1)=-\frac{1}{2}\sum_{k=1}^{\infty}A_{k}\alpha_{k}(1+\alpha_{k})v^{\alpha_{k}-1},
\end{equation}
where we have used the property
\begin{equation*}
\lim_{q\rightarrow 0}\frac{P_{\alpha}^{1}(1-2q)}{2\sqrt{q(1-q)}}=-\frac{1}{2}\alpha(1+\alpha).
\end{equation*}
Similarly, using Eq.~(\ref{xlessl}), we obtain
\begin{equation}\label{xq0l}
x(q=0,0 \leq v \leq 1)=-\sum_{k=1}^{\infty}[A_{k}(v^{\alpha_{k}}-1)(\alpha_{k}^{2}-1)+k(2+k)C_{k}].
\end{equation}
In their turn, Eqs. (\ref{tmore}) and (\ref{xmore}) yield at $q=0$:
\begin{eqnarray}
  t(q=0,v \geq 1)&=& 1-\frac{1}{2}\sum_{l=1}^{\infty}\,l(l+1)C_{l} v^{-l-2}, \label{tq0m}\\
  x(q=0,v \geq 1) &=& -\sum_{l=1}^{\infty}l(l+2)C_{l}v^{-l-1}. \label{xq0m}
\end{eqnarray}
Now, using Eq.~(\ref{Hopfvoidsol}), we can calculate $F_{\text{void}}(v)=x(0,v)-2vt(0,v)$. We obtain
\begin{equation}
F_{\text{void}}(v)=\left\{
\begin{array}{lcc}
\sum_{k=1}^{\infty}[(1+\alpha_{k})A_{k}(v^{\alpha_{k}}+\alpha_{k}-1)+k(k+2)C_{k}],& \mbox{~~for~~} & 0\leq v \leq 1,\\
-2v-\sum_{k=1}^{\infty}kC_{k}v^{-k-1},& \mbox{~~for~~} & v \geq 1.
\end{array}
\right.
\label{Fvoid}
\end{equation}
The plot of function $F_{\text{void}}(v)$ in the particular case of $n_0=1/4$ is shown
in Figure \ref{FcustrFvoid} (left panel). Note that,
at fixed $t$, there are two branches of the solution for $v$ versus $x$. It is the upper branch (the one with greater values of $v$) which should be chosen, whereas the lower branch (the one with smaller values of $v$) of $v(x,t)$ must be discarded. This is because $v$ vanishes on the lower branch, at all times,  at the point $x=x_-$, and grows with $x$ monotonically on the interval $x_-<x<x_0(t)$. As a result, the lower-branch solution for $v$ does not exhibit shock discontinuity. Furthermore, as $t\to 1$, this solution remains non-zero on the whole interval $x_-\leq x<0$ and does not obey the boundary condition  $v(x,t=1)=\delta(x)$.

\begin{figure}
\includegraphics[scale=0.5] {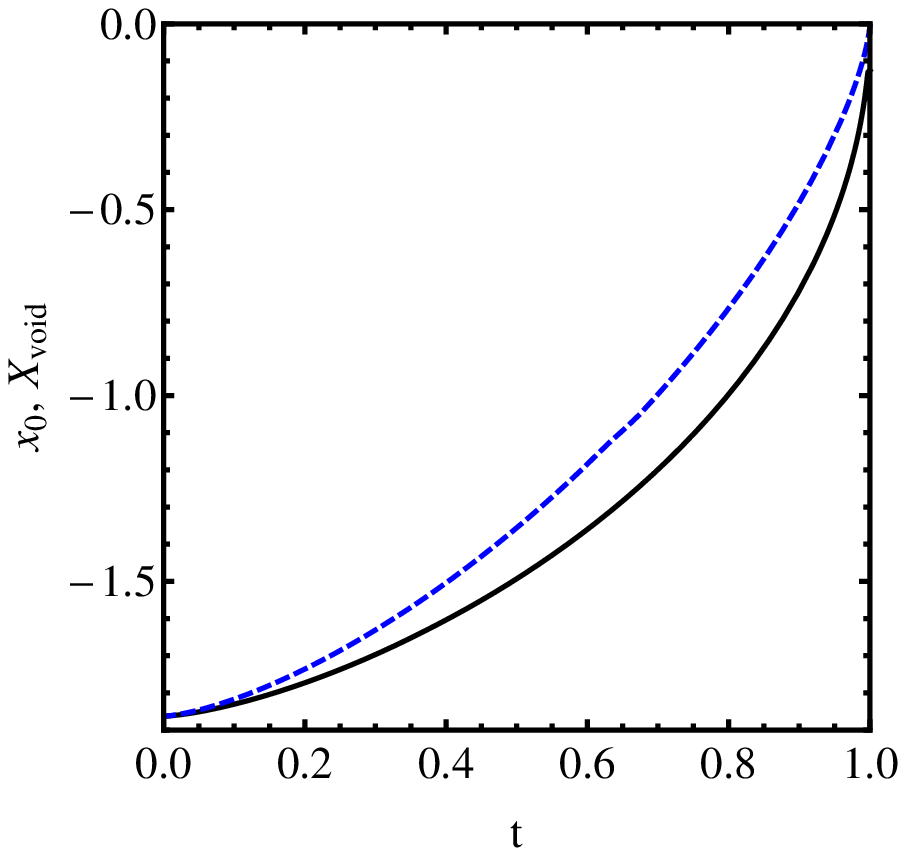}
\includegraphics[scale=0.48] {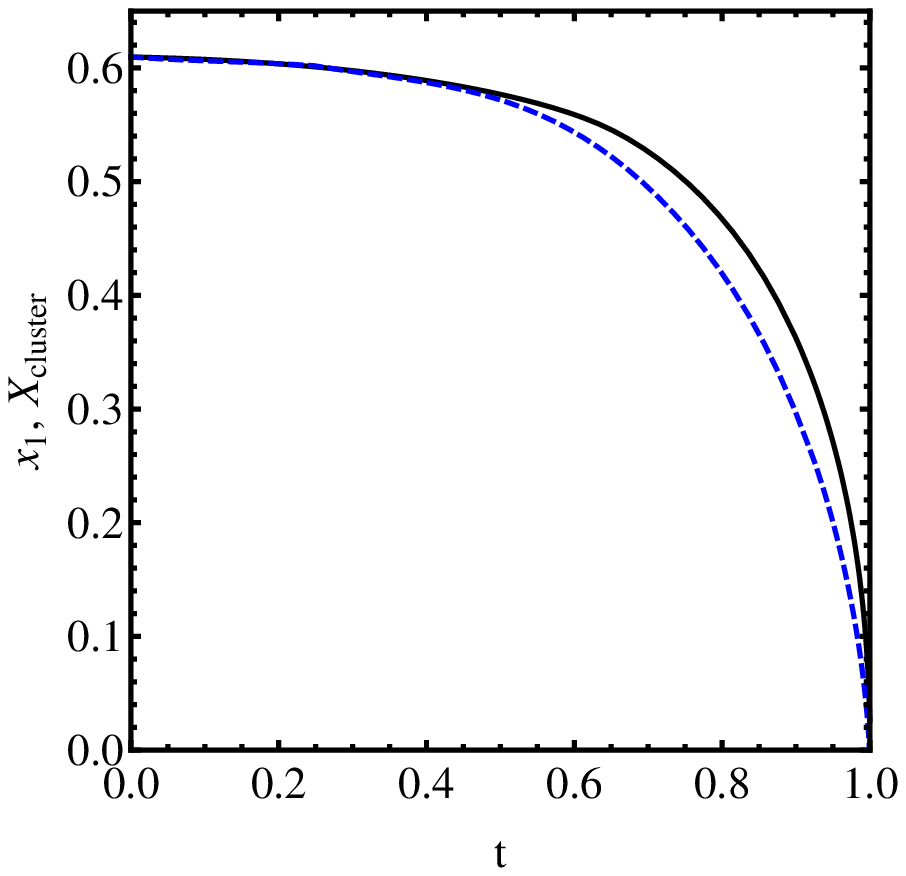}
\caption{Left panel: the border point $x_0(t)$ between the $0<q<1$ region and the void region $q=0$ (dashed line), and the coordinate $X_{\text{void}}(t)$ of the moving shock in the void region (solid line) are shown as functions of time.  Right panel: the border point $x_1(t)$ between the  $0<q<1$ region and the close-packed cluster region $q=1$ (dashed line) and the coordinate $X_{\text{cluster}}(t)$ of the moving shock in the cluster region (solid line) are shown as functions of time. The density $n_0=1/4$. The parametrization $v_{\text{max}}(x,t=0)=1$ is used.}
\label{x0Xvx1Xc}
\end{figure}

The upper-branch solution $v(x,t)$ in the void region holds for $X_{\text{void}}(t)<x<x_0(t)$. Here $x_0(t)$ is the (time-dependent) ``left" edge of the hodograph region,
determined in parametric form by Eqs.~(\ref{tq0l}) and (\ref{xq0l}) for $0\leq v\leq 1$ and by Eqs.~(\ref{tq0m}) and (\ref{xq0m}) for $v\geq 1$. In its turn, $X_{\text{void}}(t)$ is the coordinate
of a shock discontinuity of $v(x,t)$ which separates the void region $x_-<x<x_0(t)$ into two subregions: $q=0,\,v=0$ at $x_-<x<X_{\text{void}}(t)$ and  $q=0,\,v=v(x,t)$ at $X_{\text{void}}(t)<x<x_0(t)$. The shock coordinate $X_{\text{void}}(t)$  is governed by Eq.~(\ref{shockvoid}). The first-order equation~(\ref{shockvoid}) should be solved with the initial condition $X_{\text{void}}=x_-$.  The solution can be obtained numerically for a given $n_0$. The quantities $x_0$ and $X_{\text{void}}$ versus time are shown in Fig. \ref{x0Xvx1Xc} (left panel) by the dashed and solid lines, respectively, for $n_0=1/4$. The resulting dynamics of $v(x,t)$ in the void region is depicted, for $n_0=1/4$, in Fig.~\ref{qvofx}.

\subsection{Cluster region}
In order to find $F_{\text{cluster}}(v)$, we need to evaluate the hodograph solution for $t(1,v)$ and $x(1,v)$. The calculations are similar
to those for the void, and we obtain
\begin{equation}
F_{\text{cluster}}(v)=x(1,v)+2vt(1,v)=\left\{
\begin{array}{lcc}
\sum_{k=1}^{\infty}[(1+\beta_{k})B_{k}(1-v^{\beta_{k}}-\beta_{k})+(-1)^{k+1}k(k+2)C_{k}]& \mbox{~~for~~} &0 \leq v \leq 1,\\
2v+\sum_{k=1}^{\infty}(-1)^{k+1}kC_{k}v^{-k-1}& \mbox{~~for~~} & v \geq 1.
\end{array}
\right.
\label{Fcluster}
\end{equation}
The plot of function $F_{\text{cluster}}(v)$ for $n_0=1/4$ are shown
in Fig.~\ref{FcustrFvoid} (right panel).

The solution $v(x,t)$ in the cluster region holds for $x_1(t)<x<X_{\text{cluster}}(t)$. Here $x_1(t)$ is the ``right" edge of the hodograph region which is obtained in parametric form by going to the limit of $q\to 1$ in Eqs (\ref{tmore}), (\ref{xmore}), (\ref{tlessm}) and (\ref{xmr}):

\begin{equation}
t(q=1,v)=\left\{
\begin{array}{lcc}
-\frac{1}{2}\sum_{k=1}^{\infty}\beta_{k}(\beta_{k}+1)B_{k}v^{\beta_{k}-1}& \mbox{~~for~~} & 0\leq v\leq 1,\\
1-\frac{1}{2}\sum_{l=1}^{\infty}\,(-1)^{l-1}l(l+1)C_{l} v^{-l-2}& \mbox{~~for~~} & v\geq 1,
\end{array}
\right.
\label{tq1}
\end{equation}
\begin{equation}
x(q=1,v)=\left\{
\begin{array}{lcc}
\sum_{k=1}^{\infty}[(\beta_{k}^{2}-1)(v^{\beta_{k}}-1)B_{k}+(-1)^{k+1}k(k+2)C_{k}]& \mbox{~~for~~} & 0\leq v\leq 1\\
\sum_{k=1}^{\infty}(-1)^{k+1}k(k+2)C_{k}v^{-k-1}& \mbox{~~for~~} & v\geq 1.
\end{array}
\right.
\label{xq1}
\end{equation}
In its turn, $X_{\text{cluster}}(t)$ is the coordinate
of a shock discontinuity of $v(x,t)$ which separates the cluster region $x_1(t)<x<x_+$ into two subregions: $q=1,\,v=0$ at $X_{\text{cluster}}(t)<x<x_+$ and  $q=1,\,v=v(x,t)$ at $x_1(t) <x<X_{\text{cluster}}(t)$. The shock coordinate $X_{\text{cluster}}(t)$  is governed by Eq.~(\ref{shockcluster}) with the initial condition $X_{\text{cluster}}(t=0)=x_+$. The quantities $x_1$ and $X_{\text{cluster}}$ versus time, for $n_0=1/4$, are shown in the right panel of Fig.~\ref{x0Xvx1Xc} by the dashed and solid lines, respectively. The resulting dynamics of $v(x,t)$ in the cluster region is shown, for $n_0=1/4$, in Fig.~\ref{qvofx}.

In the particular case of $n_0=1/2$ the hodograph solution is elementary \cite{MS2014}, and one obtains
\begin{equation}\label{Fsimple}
   F_{\text{cluster}}(v)=- F_{\text{void}}(v)=\frac{4}{\pi} \left(1+v \arctan v \right).
\end{equation}
In this case the whole solution is symmetric (that is, $v$ is an even function of $x$, and $q-1/2$ is an odd function of $x$) at all times.

\subsection{Void and cluster at $t\to 1$}

What happens when $t$ approaches 1?  Here $x_1(t)\simeq -x_0(t)\simeq 3\ell(t)= 3C_1^{1/3} (1-t)^{2/3}$, see Eq.~(\ref{similarityq}).  Furthermore, as  $v$ is very large, we can only account for the leading and subleading terms in the second lines of Eq.~(\ref{Fvoid}) and (\ref{Fcluster}). This yields the asymptotic
\begin{equation}\label{Hopfasymp}
    |x| \simeq 2 (1-t) v+\frac{C_1}{v^2},
\end{equation}
or
\begin{equation}\label{Hopfasymp1}
    \frac{|x|}{\ell(t)} \simeq 2 V+\frac{1}{V^2},\;\;\;\;\text{where}\;\;\;\;V=C_1^{1/3} (1-t)^{1/3} v.
\end{equation}
That it, the Hopf flows become self-similar with the same similarity indices as in the region of $|x|<3 \ell (t)$, where $0<q<1$.  As we noticed earlier, see Eq.~(\ref{similarityv}), $v(x,t)$ in the region of $0<q<1$ develops a plateau as $t \to 1$. Now we see from Eq.~(\ref{Hopfasymp}) that $v(x,t\to 1)$ is symmetric with respect to $x$ in the Hopf flow regions. Therefore, $v$ becomes an even function of $x$ \emph{everywhere} as $t \to 1$.

There is, however, an important additional dynamic length scale in the Hopf regions. This length scale is introduced by the shock positions $X_{\text{void}}(t)$ and $X_{\text{cluster}}(t)$ which are governed by Eqs.~(\ref{shockvoid}) and (\ref{shockcluster}). At $t\to 1$, we can drop the $C_1/v^2$ term in Eq.~(\ref{Hopfasymp}) in most of the Hopf flow regions, except very close to the hodograph region. This yields $X_{\text{cluster}}(t) \simeq -X_{\text{void}}(t) \equiv X(t) \simeq A(1-t)^{1/2}$, where the coefficient $A$ will be found shortly. The maximum values of $v$ are reached at the shocks, $x=\pm X(t)$. Here $v[\pm X(t), t]=v_{\text{max}}(t)\simeq (A/2) (1-t)^{-1/2}$.

The length scale $X(t)\sim (1-t)^{1/2}$ is much greater than $\ell(t)\sim (1-t)^{2/3}$, and $v$ grows with the distance as one exits the hodograph region, reaching the maximum at $x=\pm X(t)$. As a result, at $t \to 1$, the two Hopf flow regions make a dominant contribution to the conservation of $v$, while the $0<q<1$ region only contributes sub-dominantly. This enables us to find the coefficient $A$ from the condition
\begin{equation}\label{findA}
   \int_{-X(t)}^{X(t)} v(x,t) \,dx \simeq \Lambda_*.
\end{equation}
A dominant contribution to the integral comes from the regions where the $C_1/v^2$ term in Eq.~(\ref{Hopfasymp}) can be dropped. Then Eq.~(\ref{findA}) yields $A\simeq \sqrt{2 \Lambda_*}$, where $\Lambda_*$ is given by Eq.~(\ref{Lambda*1}). We emphasize that, in contrast to the contribution to $\Lambda_*$,  \emph{all} of the contribution to the action $s_*$
comes from the $0<q<1$ region, at all times.

\section{Discussion}
\label{discus}

We employed the macroscopic fluctuation theory (MFT) to evaluate the logarithm of the probability distribution $\ln {\cal P}\simeq -\Phi(n_0) J^3/T$ of observing very large values of integrated current $J$ at time $T$ for the SSEP, when starting from a flat density profile $n=n_0$, with any $0<n_0<1$, on an infinite line. We calculated the function $\Phi(n_0)$ analytically. We found that $\Phi(n_0)$ exhibits a singularity $\Phi(n_0) \sim |n_0-1/2|$ at the half-filling density, see Fig.~\ref{Phifig}. We argue that this singularity only appears in the zero-diffusion approximation, whereas the exact large deviation function of current $s(j,n_0,n_0)$ is smooth inside a narrow boundary layer around the half-filling density.

Although we did not attempt to calculate the subleading correction to our main result (\ref{result}), we expect
that it will be smaller by a factor $1/j=\sqrt{T}/J \ll 1$. The accuracy of our
results for the optimal path, in its different regions, is a more complicated issue which is beyond this paper.

We have been able to calculate $\Phi(n_0)$ because we exactly solved the MFT equations in the inviscid limit. The solution yields the optimal path of the system: the most probable history of the system conditional on the extreme current. In this non-stationary setting, the optimal path turns out to be surprisingly complicated, see Fig.~\ref{qvofx}.  It includes 7 different regions of smooth flow, separated by static and traveling discontinuities. These discontinuities become narrow boundary layers when finite diffusion is reintroduced. Alongside with
$\Phi(n_0)$, the (extreme-current limit of the) optimal path is a central result of this paper.  With some work, this result can be extended to a general step-like initial condition, $n_-\neq n_+$.

On a more general note, the MFT (and other similar classical field theories, such as the  celebrated Martin-Siggia-Rose theory \cite{MSR} for continuous stochastic systems) proved to be invaluable tools for studying large deviations in nonequilbrium stochastic systems. The MFT equations -- coupled nonlinear partial differential equations -- are usually hard to solve analytically, unless one can figure out the general character of the solution or even guess the correct ansatz. The  list of problems which have been solved so far in the context of diffusive lattice gases is quite short, and every new solved problem (or even its extreme limit, as reported in this paper and Refs. \cite{MS2013,MS2014}) is important.

\section*{Acknowledgments}
We are grateful to P. L. Krapivsky for useful discussions. AV and BM were supported by the US-Israel
Binational Science Foundation, grant No. 2012145. PVS was supported by the Russian
Foundation for Basic Research, grant No. 13-01-00314.

\appendix

\section{Hodograph Transformation}
\label{hodograph}

To go over from $q(x,t)$ and $v(x,t)$ to $x(q,v)$ and $t(q,v)$, we use the following identities:
\begin{eqnarray}
  \left(\frac{\partial x}{\partial t}\right)_x &=& \partial_q x \,\partial_t q+ \partial_v x \, \partial_t v=0, \label{xt}\\
  \left(\frac{\partial t}{\partial t}\right)_x &=& \partial_q t \,\partial_t q+ \partial_v t \,\partial_t v=1, \label{tt}\\
   \left(\frac{\partial x}{\partial x}\right)_t &=& \partial_q x \,\partial_x q+ \partial_v x \,\partial_x v=1, \label{xx}\\
    \left(\frac{\partial t}{\partial x}\right)_t &=& \partial_q t \,\partial_x q+ \partial_v t \,\partial_x v=0. \label{tx}
\end{eqnarray}
Equations~(\ref{xt}) and (\ref{tt}) yield $\partial_t q=-\partial_v x/{\cal J}$ and
$\partial_t v=\partial_q x/{\cal J}$, where
\begin{equation}\label{calJ}
{\cal J}=\frac{\partial(x,t)}{\partial(q,v)} = \sigma(q) (\partial_q t)^2 -\frac{1}{2} \sigma^{\prime\prime}(q) v^2 (\partial_v t)^2 = \frac{1}{\sigma(q) (\partial_x v)^2-(1/2) \sigma^{\prime\prime}(q) v^2 (\partial_x q)^2}
\end{equation}
is the Jacobian of hodograph transformation. In their turn, Eqs.~(\ref{xx}) and (\ref{tx}) yield $\partial_x q=-\partial_v t/{\cal J}$ and $\partial_x v=-\partial_q t/{\cal J}$.
Plugging these four expressions in Eqs.~(\ref{d1}) and (\ref{d2}) we arrive at Eqs.~(\ref{Ahod}) and (\ref{Bhod}).
For the SSEP we have
\begin{equation}\label{calJSSEP}
{\cal J}=2 q(1-q)(\partial_q t)^2+2 v^2 (\partial_v t)^2 = \frac{1}{2 q (1-q) (\partial_x v)^2 +  2 v^2 (\partial_x q)^2}.
\end{equation}

\section{Effective charge density}
\label{sigmacoeff}
Using Eqs.~(\ref{Psi1}) and~(\ref{10}), we obtain the following equation for the unknown coefficients $c_k$
in the expansion (\ref{sigma}) of the effective surface charge density:
\begin{equation}
\sum_{k=1}^{\infty}c_{k}\int_{0}^{2\pi}d\xi\cos\xi\int_{0}^{1}\frac{dr^{\prime}(r^{\prime})^{\gamma_{k}}}
{\sqrt{r^{2}+r^{\prime 2}-2rr^{\prime}\cos\Gamma(\theta_{0},\xi)}}=r,
  \label{main}
\end{equation}
where 
\begin{equation}\label{cosg}
\cos\Gamma(\theta_{0},\xi)=\cos^{2}\theta_{0}+\sin^{2}\theta_{0}\cos\xi .
\end{equation}
We evaluate the integral over $r^{\prime}$, using the fact that 
\begin{equation*}
\frac{1}{\sqrt{r^{2}+r^{\prime 2}-2rr^{\prime}\cos\Gamma(\theta_{0},\xi)}}
\end{equation*}
is the generating function of Legendre polynomials \cite{Jackson} and dividing the
integration domain into two domains, $0<r^{\prime}<r$ and $r<r^{\prime}<1$:
\begin{eqnarray}
 \int_{0}^{1}\frac{dr^{\prime}(r^{\prime})^{\gamma_{k}}}
 {\sqrt{r^{2}+r^{\prime 2}-2rr^{\prime}\cos\Gamma(\theta_{0},\xi)}} &=&
 \sum_{l=0}^{\infty}\int_{0}^{r}P_{l}[\cos\Gamma(\theta_{0},\xi)]
 \frac{(r^{\prime})^{\gamma_{k}+l}}{r^{l+1}}dr^{\prime}+
\sum_{l=0}^{\infty}\int_{r}^{1}P_{l}[\cos\Gamma(\theta_{0},\xi)]
\frac{(r^{\prime})^{\gamma_{k}}r^{l}}{(r^{\prime})^{l+1}}dr^{\prime}
\nonumber \\
 &=& \sum_{l=0}^{\infty}
 \left(\frac{r^{\gamma_{k}}}{\gamma_{k}+l+1}-\frac{r^{\gamma_{k}}}{\gamma_{k}-l}+\frac{r^{l}}{\gamma_{k}-l}\right)\,
 P_{l}[\cos\Gamma(\theta_{0},\xi)],
 \label{int1}
\end{eqnarray}
where $P_{l}(\cos \Gamma)$ are the Legendre polynomials. Now we perform integration over $\xi$ in Eq.~(\ref{main}), using Eq.~(\ref{int1}). 
First, we will show that
\begin{equation*}
 \int_{0}^{2\pi}d\xi \cos \xi \sum_{l=0}^{\infty}
 \left(\frac{r^{\gamma_{k}}}{\gamma_{k}+l+1}-\frac{r^{\gamma_{k}}}{\gamma_{k}-l}\right) \,P_{l}[\cos\Gamma(\theta_{0},\xi)]= 0.
\end{equation*}
This boils down to showing that
\begin{equation}\label{aux1}
\int_{0}^{2\pi}d\xi \cos \xi \sum_{l=0}^{\infty} \frac{2l+1}{(\gamma_{k}-l)(\gamma_{k}+l+1)}\,P_{l}[\cos\Gamma(\theta_{0},\xi)]=0.
\end{equation}
Using the identity (see http://dlmf.nist.gov/14.18.E8)
\begin{equation*}\sum_{l=0}^{\infty}\frac{2l+1}{(\nu-l)
(\nu+l+1)}P_{l}(z) =\frac{\pi\, P_{\nu}(-z)}{\sin(\nu\pi)},
\end{equation*}
we can simplify the left hand side of Eq. (\ref{aux1}):
\begin{equation*}
\int_{0}^{2\pi}d\xi \cos \xi \sum_{l=0}^{\infty} \frac{2l+1}{(\gamma_{k}+l+1)(\gamma_{k}-l)}\,P_{l}[\cos\Gamma(\theta_{0},\xi)]=
\frac{\pi}{\sin(\pi\gamma_{k})}\int_{0}^{2\pi}d\xi \cos \xi P_{\gamma_{k}}[-\cos\Gamma(\theta_{0},\xi)],
\end{equation*}
where $\cos \Gamma (\theta_0,\xi)$ is defined in Eq.~(\ref{cosg}). 
In the last integral we use the following identity (see http://dlmf.nist.gov/14.18.E1):
\begin{equation*}
P_{\nu}(\cos\theta_{1}\cos\theta_{2}+\sin\theta_{1}\sin\theta_{2}\cos\phi)=P_{\nu}(\cos\theta_{1})P_{\nu}(\cos\theta_{
2})+2\sum_{m=1}^{\infty}(-1)^{m}P^{-m}_{\nu}(\cos\theta_{1})P^{
m}_{\nu}(\cos\theta_{2})\cos(m\phi),
\end{equation*}
and choose $\theta_{1}=\theta_{0}$ and $\theta_{2}=\pi+\theta_{0}$. We obtain
\begin{eqnarray}
\nonumber
&&\frac{\pi}{\sin(\pi\gamma_{k})}\int_{0}^{2\pi}d\xi \cos \xi P_{\gamma_{k}}[-\cos\Gamma(\theta_{0},\xi)]  \\ \nonumber
&=& \frac{\pi}{\sin(\pi\gamma_{k})}\int_{0}^{2\pi}d\xi \cos \xi\left[ P_{\gamma_{k}}(\cos\theta_{0})P_{\gamma_{k}}(-\cos\theta_{0})
+2\sum_{m=1}^{\infty}(-1)^{m}P^{-m}_{\gamma_{k}}(\cos\theta_{0})P^{
m}_{\gamma_{k}}(-\cos \theta_{0})\cos(m\xi)\right] \\ \nonumber
&=&-\frac{2\pi^2}{\sin(\pi\gamma_{k})} P^{-1}_{\gamma_{k}}(\cos\theta_{0})P^{
1}_{\gamma_{k}}(-\cos\theta_{0})=0,
\end{eqnarray}
because $P^{1}_{\alpha_{k}}(\cos\theta_{0})=0$, $P^{1}_{\beta_{k}}(-\cos \theta_{0})=0$, and
\begin{equation}\label{lgndridentity2}
P^{-1}_{\nu}(z)=- \frac{\Gamma(\nu)}{\Gamma(\nu+2)}P^{1}_{\nu}(z),
\end{equation}
see http://dlmf.nist.gov/14.9.E3. This verifies Eq. (\ref{aux1}), and the double integral in Eq. (\ref{main}) becomes:
\begin{equation*}
\int_{0}^{2\pi} d\xi\cos\xi\sum_{l=0}^{\infty}P_{l}[\cos\Gamma(\theta_{0},\xi)]
 \frac{r^{l}}{\gamma_{k}-l}.
\end{equation*}
To evaluate the integral
\begin{equation}\label{identity}
\int_{0}^{2\pi}d\xi \cos\xi P_{l}[\cos\Gamma(\theta_{0},\xi)],
\end{equation}
we employ the identity (http://dlmf.nist.gov/14.18.E2)
\begin{equation*}
P_{l}(\cos^{2}\theta_{0}+\sin^{2}\theta_{0}\cos\xi)=\sum_{m=-l}^{
l}(-1)^{m}P^{-m}_{l}(\cos\theta_{0})P^{m}_{l}(\cos\theta_{0})\cos(m\xi)
\end{equation*}
and obtain
\begin{equation*}
\int_{0}^{2\pi}d\xi \cos\xi P_{l}[\cos\Gamma(\theta_{0},\xi)]=-\pi P^{-1}_{l}(\cos\theta_{0})P^{1}_{l}(\cos\theta_{0})=
\frac{2\pi}{l(l+1)}[P_{l}^{1}(\cos\theta_{0})]^{2}.
\end{equation*}
The latter equality stems from $P^{-1}_{l}(\cos\theta_{0})=-P^{1}_{l}(\cos\theta_{0})/[l(l+1)]$, see Eq. (\ref{lgndridentity2}), whereas the only non-vanishing terms in the sum are $m=\pm 1$. As a result,  Eq.~(\ref{main}) becomes
\begin{equation}\label{summ1}
   \sum_{k=1}^{\infty}c_{k}
\sum_{l=1}^{\infty}g^{k}_{l}(\theta_{0})\, r^{l} =r,\;\;\;\text{where}\;\;\;g^{k}_{l}(\theta_{0})=\frac{2\pi [P_{l}^{1}(\cos\theta_{0})]^{2}}{l(l+1)(\gamma_{k}-l)}.
\end{equation}
Equation~(\ref{summ1}) yields an infinite set of linear algebraic equations
\begin{equation}\label{algebraic}
\sum_{k=1}^{\infty}g^{k}_{l}c_{k} =\delta_{l1},
\end{equation}
where $\delta_{ij}$ is the Kronecker delta. Canceling the $k$-independent factors,  we arrive at a matrix equation
\begin{equation}\label{mtrxeq}
G \,\mathbf{c} = \mathbf{f},
\end{equation}
with an infinite matrix
$$
G =
 \begin{pmatrix}
\frac{1}{\gamma_1-1} & \frac{1}{\gamma_2-1} & \cdots &\frac{1}{\gamma_k-1} & \cdots \\
\frac{1}{\gamma_1-2} & \frac{1}{\gamma_2-2} & \cdots & \frac{1}{\gamma_k-2} & \cdots \\
\frac{1}{\gamma_1-3} & \frac{1}{\gamma_2-3} & \cdots & \frac{1}{\gamma_k-3} & \cdots \\
\vdots & \vdots & \vdots & \vdots & \vdots
 \end{pmatrix}
$$
and an infinite vector
$\mathbf{f}=(\pi^{-2} \sin^{-2} \theta_{0},0,0, \dots)$. One can  prove by induction that the solution of a \emph{truncated} version of Eq.~(\ref{mtrxeq}), containing
$m$ equations ($1\leq k\leq m)$, is the following:
\begin{equation}\label{truncatedsol}
    c_{k}^{\text{truncated}}=\frac{\gamma_{1}-1}{\pi\sin^{2}\theta_{0}}
\left(
 \prod_{l=1}^{m}\!\!\!\!~^{~^\prime}  \frac{1}{\gamma_{k}-\gamma_{l}}\right)
\prod_{l=2}^{m}\frac{(\gamma_{k}-l)(\gamma_{l}-1)}{l-1},
\end{equation}
where the symbol ``$~^\prime~$'' in the product shows that the multiplier with $k=l$ is skipped.
Sending $m$ to infinity, we obtain
\begin{equation}\label{c_n solution}
c_{k}=\frac{(\gamma_{1}-1)(\gamma_{k}-k-1)(\gamma_{k+1}-1)}{\pi k\sin^{2}\theta_{0}}
\prod_{l=1}^{\infty}\!\!\!\!~^{~^\prime}  \frac{(\gamma_{k}-l-1)(\gamma_{l+1}-1)}{l(\gamma_{k}-\gamma_{l})}.
\end{equation}
To remind the reader, the coefficients $c_k$ represent the union of the coefficients $a_{k}$ and $b_k$ in Eq. (\ref{sigma}). The explicit formulas for $a_k$ and $b_k$ are the following:
\begin{equation} \label{akm}
a_{k}=\frac{(\alpha_{k}-1)(\beta_{k}-1)(\alpha_{k}-k-1)}{\pi k(\alpha_{k}-\beta_{k})\sin^{2}\theta_{0}} \prod_{l=1}^{\infty}\!\!\!\!~^{~^\prime}\frac{(\alpha_{l}-1)(\beta_{l}-1)(\alpha_{k}-l-1)}{l(\alpha_{k}-\alpha_{l})(\alpha_{k}-\beta_{l})},
\end{equation}
\begin{equation} \label{bkm}
b_{k}=\frac{(\alpha_{k}-1)(\beta_{k}-1)(\beta_{k}-k-1)}{\pi k(\beta_{k}-\alpha_{k})\sin^{2}\theta_{0}} \prod_{l=1}^{\infty}\!\!\!\!~^{~^\prime}\frac{(\alpha_{l}-1)(\beta_{l}-1)(\beta_{k}-l-1)}{l(\beta_{k}-\beta_{l})(\beta_{k}-\alpha_{l})},
\end{equation}

\section{Calculating  $C_{k}$}
\label{Psicoeff}

For $r>1$ (region 3) the integral  in Eq. (\ref{10}) can be evaluated as follows:
\begin{equation*}
\int_{0}^{1}\frac{dr^{\prime}(r^{\prime})^{\gamma_{k}}}{\sqrt{r^{2}+r^{\prime 2} -2rr^{\prime}\cos\Gamma(\theta_{0},\theta,\xi)}}=
\sum_{l=0}^{\infty}\int_{0}^{1}\frac{(r^{\prime})^{\gamma_{k}+l}}{r^{1+l}} P_{l}(\cos \Gamma) dr^{\prime}=
\sum_{l=0}^{\infty}\frac{P_{l}(\cos \Gamma )}{\gamma_{k}+l+1}r^{-l-1}.
\end{equation*}
Then
\begin{equation*}
\Psi(r>1,\theta,\phi)=-\cos\phi\sin\theta_{0}\sum_{k=1}^{\infty}c_{k}
\left(\sum_{l=0}^{\infty}\frac{1}{\gamma_{k}+l+1}r^{-l-1}\int_{0}^{2\pi} P_{l}(\cos\theta\cos\theta_{0}+\sin\theta\sin\theta_{0}\cos\xi)\cos\xi d\xi\right).
\end{equation*}
The integral over $\xi$ can be evaluated:
\begin{equation}\label{int10}
\int_{0}^{2\pi} P_{l}(\cos\theta\cos\theta_{0}+\sin\theta\sin\theta_{0}\cos\xi)\cos\xi d\xi=\frac{2\pi}{l(l+1)}P_{l}^{1}(\cos\theta)P_{l}^{1}(\cos\theta_{0}),\;\;\;l=1,2, \dots ;
\end{equation}
it vanishes for $l=0$. As a result,
\begin{equation}\label{psioutr}
\Psi(r>1,\theta,\phi)=-2\pi\cos\phi\sin\theta_{0}\sum_{l=1}^{\infty}\frac{P_{l}^{1}(\cos\theta)P_{l}^{1}(\cos\theta_{0})}{l(l+1)} \left(\sum_{k=1}^{\infty}\frac{c_{k}}{\gamma_{k}+l+1}\right)r^{-l-1}.
\end{equation}
Comparing this expression with Eq.~(\ref{region3}), we obtain
\begin{equation}\label{C_l}
C_{l}=-2\pi\sin\theta_{0}\frac{P^{1}_{l}(\cos\theta_{0})}{l(l+1)}\sum_{k=1}^{\infty}
\frac{c_{k}}{\gamma_{k}+l+1},
\end{equation}
which yields Eq.~(\ref{C_lsecond}).

\section{Finding $x(q,v)$ of the hodograph solution}
\label{xhod}
\subsection{$v \geq 1$}
\label{xhodmore}
Calculating the partial derivatives $\partial_q t$ and $\partial_v t$  of $t(q,v \geq 1)$ from Eq.~(\ref{tmore}), we obtain
\begin{eqnarray}
  \partial_q t (q,v \geq 1) &=& \frac{1}{4[(1-q)q]^{3/2}}\sum_{l=1}^{\infty}C_{l}v^{-l-2}[(2+l)(2q-1)P^{1}_{l}(1-2q)+lP^{1}_{l+1}(1-2q)], \label{tql}\\
   \partial_v t (q,v \geq 1)  &=& -\frac{1}{2\sqrt{(1-q)q}}\sum_{l=1}^{\infty}(l+2)C_{l}v^{-l-3}P^{1}_{l}(1-2q).\label{tvl}
\end{eqnarray}
Plugging these into Eq.~(\ref{xintegral}) and performing the integration, we arrive at Eq.~(\ref{xmore}).

\subsection{$0 \leq v \leq 1$}
\label{xhodless}

Here we need to calculate the partial derivatives $\partial_q t$ and $\partial_v t$ in the two subregions $0 \leq q \leq n_0$ and $n_0 \leq q \leq 1$:
\begin{eqnarray}
  \partial_qt(0 \leq q \leq n_0,0 \leq v \leq 1) &=& \frac{\sum_{k=1}^{\infty}A_{k}v^{\alpha_{k}-1}\left[(2+\alpha_{k})
(2q-1)P^{1}_{\alpha_{k}}(1-2q)+\alpha_{k}P^{1}_{\alpha_{k}+1}(1-2q)\right]}{4[q(1-q)]^{3/2}}, \label{der1}\\
  \partial_qt(n_0 \leq q \leq 1,0 \leq v \leq 1) &=& \frac{\sum_{k=1}^{\infty}B_{k}v^{\beta_{k}-1}
\left[(2+\beta_{k})(2q-1)P^{1}_{\beta_{k}}(2q-1)-\beta_{k}P^{1}_{\beta_{k}+1}(2q-1)\right]}{4[q(1-q)]^{3/2}}, \label{der2}\\
 \partial_vt(0 \leq q \leq n_0, 0 \leq v \leq 1) &=&
\frac{\sum_{k=1}^{\infty}A_{k}(\alpha_{k}-1)v^{\alpha_{k}-2}P^{1}_{\alpha_{k}}(1-2q)}{2\sqrt{q(1-q)}}, \label{der3}\\
 \partial_vt(n_0 \leq q \leq 1, 0 \leq v \leq 1) &=&
\frac{\sum_{k=1}^{\infty}B_{k}(\beta_{k}-1)v^{\beta_{k}-2}P^{1}_{\beta_{k}}(2q-1)}{2\sqrt{q(1-q)}} \label{der4}.
\end{eqnarray}
Now we rewrite Eq. (\ref{xintegral}) as a sum of two integrals:
\begin{equation}\label{xintegral2}
x(q,v )= \int_{v}^{1}\left[2(2q-1) v \partial_v t +2q(1-q)\partial_q t\right]dv +
\int_{1}^{\infty}\left[2(2q-1) v \partial_v t +2q(1-q)\partial_q t\right]dv.
\end{equation}
Plugging Eqs.~(\ref{der1})-(\ref{der4}) in the first term, and Eqs.~(\ref{tql}) and (\ref{tvl}) in the second one (separately in the two subregions of $q$), and performing the integrations, we obtain Eqs.~(\ref{xlessl}) and (\ref{xmr}).


\begin{thebibliography} {99}

\bibitem{Spohn} H. Spohn, {\it Large Scale Dynamics of Interacting Particles} (New York: Springer-Verlag, 1991).

\bibitem{L99}
    T. M. Liggett, {\it Stochastic Interacting Systems: Contact, Voter, and Exclusion Processes}
    (Springer, New York, 1999).

\bibitem{KL99}
     C. Kipnis and C. Landim, {\it Scaling Limits of Interacting Particle Systems}
     (Springer, New York,  1999).

 \bibitem{SZ95}
     B. Schmittmann and R. K. P. Zia, \textit{Statistical Mechanics of Driven Diffusive Systems},
     in: {\it Phase Transitions and Critical Phenomena}, Vol.\ 17, eds.\ C. Domb and J. L. Lebowitz
     (Academic Press, London, 1995).


\bibitem{S00}
    G.M. Sch\"utz, \textit{Exactly Solvable Models for Many-Body Systems Far From Equilibrium}, in {\it Phase Transitions and Critical Phenomena}, Vol.\ 19, eds.\ C. Domb and J. L. Lebowitz (Academic Press, London, 2000).

\bibitem{D07}
     B. Derrida, J. Stat. Mech. P07023 (2007).

\bibitem{BE07}
    R. A. Blythe and M. R. Evans, J. Phys.\ A {\bf 40}, R333 (2007).

\bibitem{KRB10}
     P. L. Krapivsky, S. Redner, and E. Ben-Naim,
     {\it A Kinetic View of Statistical Physics} (Cambridge University Press, Cambridge,  2010).

     \bibitem{Levitov} H. Lee, L.S. Levitov, and A. Yu. Yakovets,
Phys. Rev. B \textbf{51}, 4079 (1995).

\bibitem{Buttiker} Y.M. Blanter and M. B\"{u}ttiker, Phys. Rep. \textbf{336}, 1 (2000).

\bibitem{Jordan1} S. Pilgram, A.N. Jordan, E.V. Sukhorukov and M. B\"{u}ttiker,  Phys. Rev. Lett. \textbf{90}, 206801 (2003).

\bibitem{Jordan2} A.N. Jordan, E.V. Sukhorukov, and S. Pilgram, J. Math. Phys. \textbf{45}, 4386 (2004).

\bibitem{Chou}
    T. Chou, K. Mallick, and R. K. P. Zia, Rep. Prog. Phys. \textbf{74}, 116601 (2011).

\bibitem{KMP}
     C. Kipnis, C. Marchioro, and E. Presutti, J. Stat. Phys. {\textbf 27}, 65 (1982).

\bibitem{Jona}


    G. Jona-Lasinio, Prog. Theor. Phys. Suppl. {\textbf 184}, 262 (2010); J. Stat. Mech. (2014) P02004.


\bibitem{HEPG} P.I. Hurtado, C. P. Espigares, J. J. del Pozo, and P. L. Garrido, J. Stat. Phys. \textbf{154}, 214 (2014).

\bibitem{DG2009a}
   B. Derrida and A. Gerschenfeld, J. Stat. Phys. \textbf{136}, 1 (2009).

\bibitem{DG2009b}
    B. Derrida and A. Gerschenfeld, J. Stat. Phys. \textbf{137}, 978 (2009).


\bibitem{KM_var}
    P. L. Krapivsky and B. Meerson,  Phys. Rev. E \textbf{86}, 031106 (2012).

\bibitem{varadhan} S. Sethuraman and S.R.S. Varadhan, Ann. Prob. \textbf{41}, 1461 (2013).

\bibitem{MS2013}
    B. Meerson and P.V. Sasorov, J. Stat. Mech. (2013) P12011.

\bibitem{MS2014}
    B. Meerson and P.V. Sasorov, Phys. Rev. E \textbf{89}, 010101(R) (2014).

\bibitem{Bertini}
    L. Bertini, A. De Sole, D. Gabrielli, G. Jona-Lasinio, and C. Landim,
    Phys. Rev. Lett. {\textbf 87}, 040601 (2001);  J. Stat. Phys. \textbf{107}, 635 (2002);
    Phys. Rev. Lett. \textbf{94}, 030601 (2005);   J. Stat. Phys. {\textbf 123}, 237 (2006).

\bibitem{Tailleur}
    J. Tailleur, J. Kurchan, and V. Lecomte, Phys. Rev. Lett. \textbf{99}, 150602 (2007);
    J. Phys. A {\bf 41}, 505001 (2008).


\bibitem{varadhan1}  The $s \sim j^3$ decay has been recently proved
for the SSEP for more general initial density profiles \cite{varadhan}.


\bibitem{LLfluidmech} L. D. Landau and E. M. Lifshitz, {\it Fluid Mechanics} (Reed, Oxford, 2000).

\bibitem{Courant}  R. Courant and K.O. Friedrichs, \textit{Supersonic Flow and Shock Waves} (Springer, New York, 1948).

\bibitem{discontinuities} The pointlike void and cluster which appear at $t = 0$ in the inviscid
limit imply a rapid formation of deep and narrow density minimum and maximum, respectively, in the
complete MFT equations (\ref{q:eqfull}) and (\ref{p:eqfull}) which include diffusion terms. Other
discontinuities, which appear in the inviscid solution, are also regularized by finite diffusion,
except for the ``true" singularity imposed by Eq. (\ref{p_step}).

\bibitem{Whitham} G. B. Whitham, \textit{Linear and Nonlinear Waves} (Wiley, New York, 1974).

\bibitem{Sommerfeld} A. Sommerfeld, \textit{Partial Differential Equations in Physics} (Academic, New York, 1949), p. 36.

\bibitem{Trubnikov} B.A. Trubnikov and S.K. Zhdanov, Phys. Rep. \textbf{155}, 137 (1987).

\bibitem{Abramowitz} M.S. Abramowitz and I.A. Stegun,  eds. {\it Handbook of Mathematical Functions} (New York, Dover, 1972).

\bibitem{union} Here and in most of the following we assume that neither of the eigenvalues $\alpha_k$ coincides with
neither of the eigenvalues $\beta_k$. This assumption of non-degeneracy breaks down when one of $\alpha_k$ and one of $\beta_k$ become \emph{the same integer}. This happens for a (measure zero) countable set of values of $n_0$. A full degeneracy is observed at the half-filling $n_0=1/2$. Here the conical surface (see Fig. \ref{cone}) degenerates into a disk, and one obtains $\alpha_k=\beta_k=2k$ for all $k=1,2,\dots$.  In the latter case it is more convenient to use the elliptic coordinates where the solution of Laplace's equations becomes one-dimensional and can be obtained in elementary functions \cite{MS2014}.

\bibitem{Jackson} J.D. Jackson, {\it Classical Electrodynamics} (New York, Wiley, 1975), p. 92.

\bibitem{MSR}
     P. C. Martin, E. D. Siggia, and H. A. Rose, Phys. Rev. A \textbf{8}, 423 (1973).



\end{thebibliography}
\end{document}